\documentclass[aps,prl,floats,twocolumn,showpacs,superscriptaddress]{revtex4-2}

\usepackage[justification=raggedright,singlelinecheck=false]{subcaption}
\usepackage[justification=raggedright,singlelinecheck=false]{caption}
\usepackage{float}
\usepackage{booktabs}
\usepackage{times}
\usepackage{graphicx}
\usepackage{graphics}
\usepackage{dcolumn}
\usepackage{bm}
\usepackage[mathlines]{lineno}
\usepackage{ragged2e}

\usepackage{amsfonts}
\usepackage{amsmath}
\usepackage{amsthm}
\usepackage[usenames,dvipsnames]{color}
\usepackage[utf8]{inputenc}
\usepackage{comment}
\usepackage[normalem]{ulem}
\usepackage{cancel}

\newcommand{\av}[1]{\langle {#1} \rangle}

 \usepackage[hidelinks,unicode=true]{hyperref}
 \hypersetup{colorlinks=true,% Don't draw a box around links, instead color every piece of text, which is a link
 	linkcolor=blue,
 	urlcolor=blue,
 	citecolor=blue,
 	pdfhighlight=/N% When clicking a link and holding the mouse button, don't use any graphical effects - i.e. the alternative would be to behave like a "button" which is pressed within the PDF
 }%\usepackage{hyperref}

\begin{document}

\title{Nesting Controls Phase Transitions in Higher-Order Contagion}

\author{Hugo P. Maia}
%\email[E-mail adress:]{hugo.maia@ufv.br (Hugo P. Maia)}
\affiliation{Departamento de F\'{\i}sica, Universidade Federal de Vi\c{c}osa, 36570-900 Vi\c{c}osa, Minas Gerais, Brazil}
\affiliation{Institute for Biocomputation and Physics of Complex Systems (BIFI), University of Zaragoza, Zaragoza, Spain.}

\affiliation{Department of Theoretical Physics, Faculty of Sciences, University of Zaragoza, Zaragoza, Spain.}

\author{Guilherme Ferraz de Arruda}
%\email{gui.f.arruda@gmail.com (Guilherme Ferraz de Arruda)}
\affiliation{Instituto de F\'{\i}sica Gleb Wataghin, Universidade de Campinas (UNICAMP), Campinas, Brazil.} 

\author{Silvio C. Ferreira}
%\email{silviojr@ufv.br (Silvio C. Ferreira)}
\affiliation{Departamento de F\'{\i}sica, Universidade Federal de Vi\c{c}osa, 36570-900 Vi\c{c}osa, Minas Gerais, Brazil}
\affiliation{Instituto de Ci\^{e}ncias Matem\'{a}ticas e de Computa\c{c}\~{a}o, Universidade de S\~{a}o Paulo, S\~{a}o Carlos, SP 13566-590, Brazil}

\author{Yamir Moreno}
%\email{yamir.moreno@gmail.com (Yamir Moreno)}

\affiliation{Institute for Biocomputation and Physics of Complex Systems (BIFI), University of Zaragoza, Zaragoza, Spain.}

\affiliation{Department of Theoretical Physics, Faculty of Sciences, University of Zaragoza, Zaragoza, Spain.}

\date{\today}

\begin{abstract}
The organization of higher-order interactions plays a central role in shaping collective dynamics, yet a general structural principle governing contagion on hypergraphs remains lacking. Here we introduce a nesting coefficient that quantifies how lower-order interactions are embedded within higher-order ones, defining a continuum between simplicial complexes and random hypergraphs. Using a higher-order susceptible-infected-susceptible model, we show that increasing nesting lowers the activation threshold and suppresses discontinuous transitions, while weak embedding favors explosive behavior. We further demonstrate that correlations between nesting and interaction order modulate the onset of activity while only weakly affecting transition discontinuity. Analysis of synthetic and empirical networks reveals that nesting strongly predicts hysteresis, establishing it as a key structural determinant of phase transitions in higher-order systems.
\end{abstract}

\maketitle

At the heart of complexity science lies the quest to understand how collective behavior emerges from patterns of interaction among system components~\cite{Dorogovtsev2022}. A paradigmatic example is contagion dynamics, whose understanding has long been dominated by models based on \emph{pairwise interactions}, typically represented using networks and graph-theoretical tools~\cite{DeArruda2018}. While this framework has been highly successful, growing evidence shows that many real-world systems are governed not only by dyadic interactions, but also by \emph{higher-order interactions} involving groups of three or more agents simultaneously~\cite{ferraz2024contagion,Bick2023}. Such interactions play a key role across diverse domains, including social conventions~\cite{Iacopini2022}, functional~\cite{Sizemore2018} and structural~\cite{Petri2014} brain networks, and collective human behavior~\cite{Battiston2025}. Because higher-order interactions can qualitatively alter system dynamics, they require theoretical frameworks that go beyond pairwise descriptions~\cite{Battiston2021,Boccaletti2023}.

Higher-order networks generalize standard networks by allowing simultaneous interactions among groups of nodes. Two main formalisms are commonly used: \textit{Simplicial Complexes} (SC) and \textit{Hypergraphs} (HG)~\cite{Bianconi2021,Bick2023}. In both cases, interactions involve $m+1$ nodes $\{i_0,i_1,\ldots,i_m\}$, referred to as $m$-simplices in SC and $m$-order hyperedges in HG. The key distinction is structural: in SCs, every higher-order interaction necessarily includes all its lower-order subsets, whereas in HGs no such constraint is imposed. As a result, SCs correspond to maximally \emph{embedded} structures, while general hypergraphs may exhibit little or no embedding across interaction orders. These structural differences have important dynamical consequences. In particular, contagion processes on SCs and HGs have been shown to display markedly different behaviors~\cite{Iacopini2019, Palafox-Castillo2022, Landry2020, ferraz2023multistability,Zhang2023,Burgio2024,Kim2023}.

Recent work has begun to explore how the embedding of lower-order interactions within higher-order ones shapes collective dynamics~\cite{Kim2023,Burgio2024,Malizia2026}. In these studies, embedding is typically controlled through simplified constructions, often restricted to triadic interactions, where overlap between pairwise and higher-order interactions can be tuned. These approaches have shown that stronger embedding facilitates contagion by lowering activation thresholds and reducing hysteresis~\cite{Kim2023,Burgio2024,Malizia2025}. However, existing frameworks are limited to specific interaction orders or rely on homogeneous assumptions, and cannot be straightforwardly extended to heterogeneous hypergraphs with arbitrary order distributions. A general, quantitative framework to characterize embedding in higher-order networks and assess its dynamical consequences remains lacking.

Here, we address this gap by introducing a \emph{nesting coefficient} that quantifies the extent to which lower-order interactions are embedded within higher-order ones, defining a structural continuum between simplicial complexes and random hypergraphs. Building on this measure, we systematically investigate how embedding shapes contagion dynamics in heterogeneous higher-order networks. Using a higher-order susceptible–infected–susceptible (SIS) model, we show that increasing embedding lowers the activation threshold and promotes continuous transitions, whereas weak embedding favors explosive (discontinuous) behavior. Crucially, we demonstrate that contagion dynamics depend not only on the average level of embedding but also on how embedding correlates with interaction order, with order-dependent patterns exerting distinct effects on activation and hysteresis. These results, validated on both synthetic and empirical networks, establish embedding as a key structural determinant of contagion dynamics in higher-order systems.

\textit{The nesting coefficient.} To quantify the degree of embedding of lower-order interactions within higher-order ones, we define the nesting coefficient, a local measure of cross-order embedding. A hyperedge $h$ of order $m_h$ has a nesting coefficient of order $m < m_h$, given by the ratio between the number of $m$-order hyperedges contained within $h$ and the maximum number of such hyperedges that can be formed among its nodes:
\begin{equation}
	C^{(m)}_h = \frac{\sum_{\{i_0,i_1, ..., i_m\} \in h} \mathbf{A}_{\{i_0,i_1, ..., i_m\}}^{(m)}}{\binom{m_h + 1}{m + 1}}.
\end{equation}
Here, $\mathbf{A}_{\{i_0,i_1, ..., i_m\}}^{(m)}$ denotes the adjacency tensor of order $m$, taking value 1 if an $m$-order hyperedge composed of the node set $\{i_0,i_1, ..., i_m\}$ exists, and 0 otherwise~\cite{Boccaletti2023,Bianconi2021}. We will use the term embedding to refer to this property in qualitative terms. By construction, $C^{(m)}_h \in [0,1]$, where $C^{(m)}_h = 0$ indicates no embedding of $m$-order interactions within $h$, and $C^{(m)}_h = 1$ indicates maximal embedding.

\begin{figure}[t!]
  \includegraphics[width=0.99\linewidth]{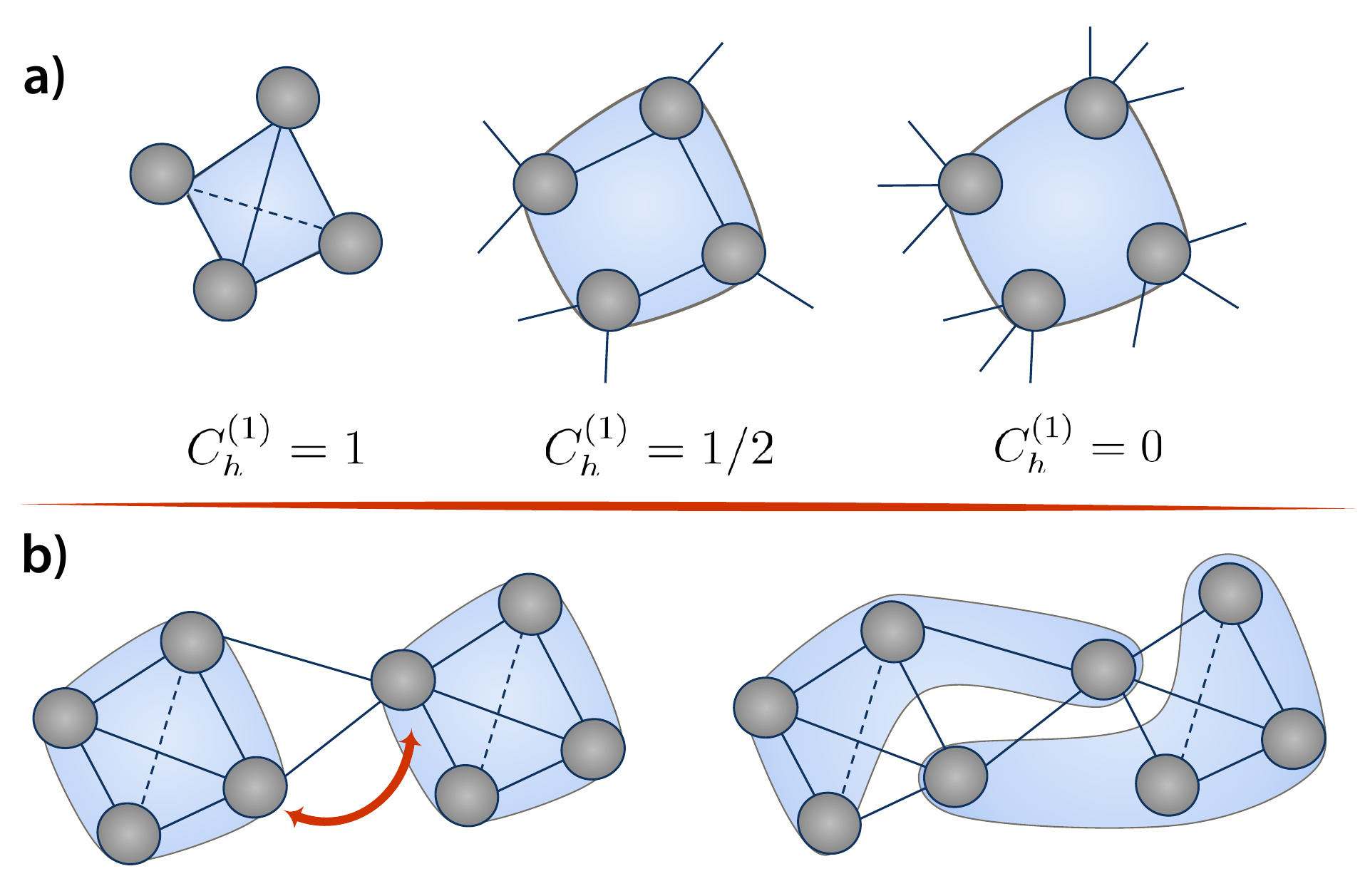}
    \caption{(a) Example of the nesting coefficient of order 1 (pairwise) in a hyperedge $h$ of order $m=3$ (4-body), $C^{(1)}_h$. The coefficient represents the ratio between the number of 1st-order hyperedges and the maximum number of hyperedges that could be formed within the hyperedge of order $m=3$. (b) Illustration of the rewiring mechanism used to control embedding. Two hyperedges of order $m=3$ exchange one node (arrow), reducing the number of embedded pairwise interactions and yielding $C^{(1)}_h = 2/3$. Simulations are performed on networks with maximum order $m_{\max}=5$ (see also SM).}
    \label{fig1}
\end{figure}

Simplicial complexes correspond to maximally embedded structures, with $C^{(m)}_h = 1$ for all $m < m_h$, since every simplex contains all its lower-order faces. In contrast, random hypergraphs lack such structural constraints, and their nesting coefficients approach zero in the sparse regime. Intermediate values arise either from high densities or, more generally, from structural correlations across interaction orders. Figure~\ref{fig1}a illustrates the nesting coefficient for hyperedges of order $m=3$.

\textit{Synthetic Higher-Order Networks.} To isolate the effects of embedding on contagion dynamics, we generate synthetic hypergraphs made up of $N$ nodes and $H$ hyperedges, controlling their nesting coefficients. Networks are constructed using an adaptation of the bipartite configuration model~\cite{Courtney2016}, preserving both interaction and group-size distributions. Starting from a simplicial complex, we apply a rewiring procedure (see Figure~\ref{fig1}b for an example) in which $f \times H$ pairs of hyperedges of the same order exchange nodes, progressively reducing the nesting coefficient. This procedure produces a continuous interpolation between maximally embedded structures ($\langle C \rangle = 1$ for $f=0$) and random hypergraphs with negligible embedding ($\langle C \rangle \to 0$ as $f \to \infty$ in the thermodynamic limit). The average nesting coefficient $\langle C \rangle$ decays approximately exponentially with the rewiring rate $f$. By introducing order-dependent rewiring biases, we further generate networks with controlled correlations between embedding and interaction order. Additional details are provided in the Supplemental Material (SM).

\begin{figure}[t!]
  \includegraphics[width=\linewidth]{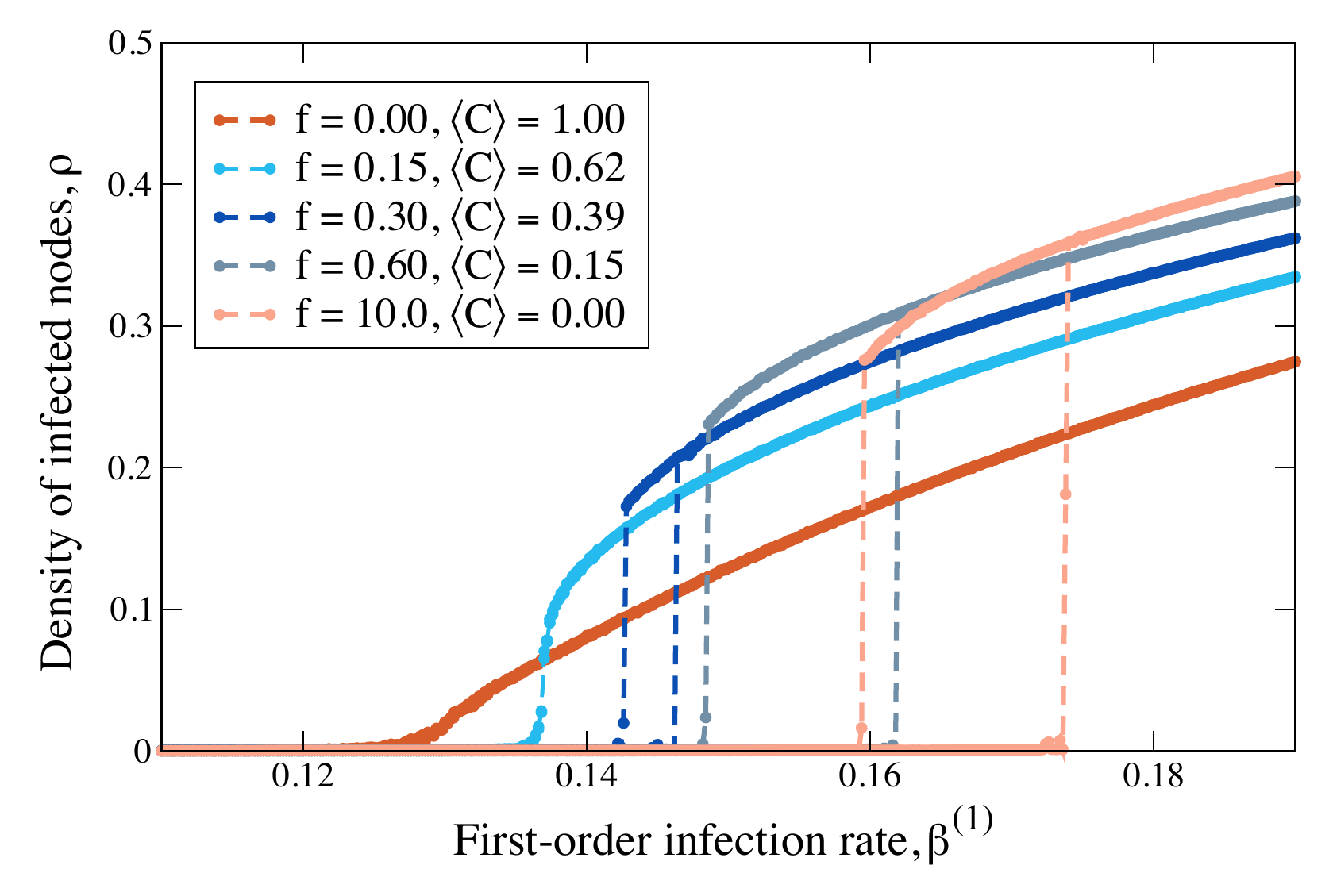}
  \caption{Stationary epidemic prevalence $\rho$ as a function of the first-order spreading rate $\beta^{(1)}$ for different rewiring levels $f$, corresponding to different average nesting coefficients $\langle C \rangle$. Forward (lower spinodal) and backward (upper spinodal) branches are obtained from initial conditions $\rho(0)\gtrsim 1/N$ and $\rho(0)\lesssim 1$, respectively.}
    \label{fig2}
\end{figure}

\begin{figure}[htbp]
         \includegraphics[width=\linewidth]{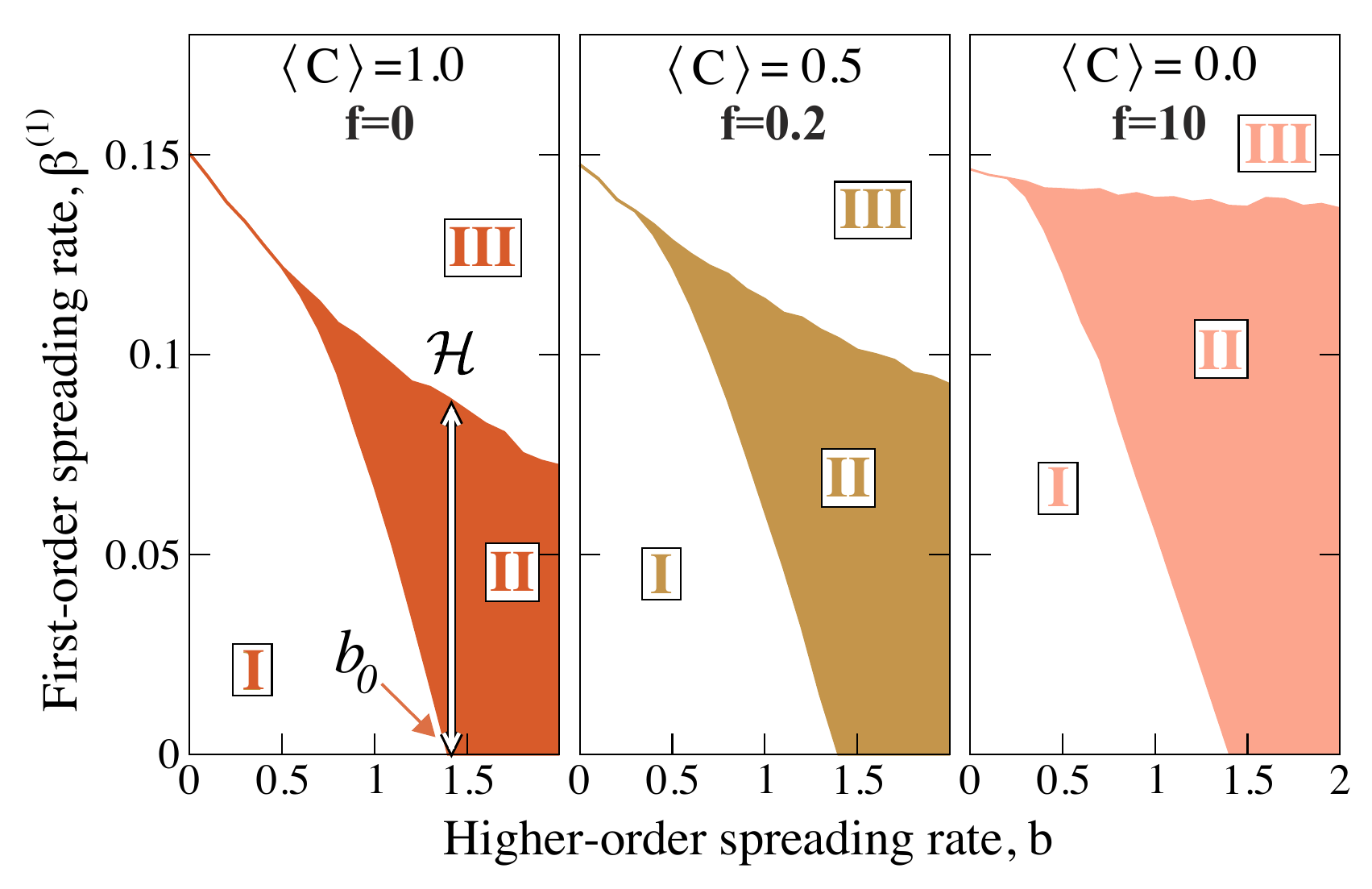}
        \includegraphics[width=\linewidth]{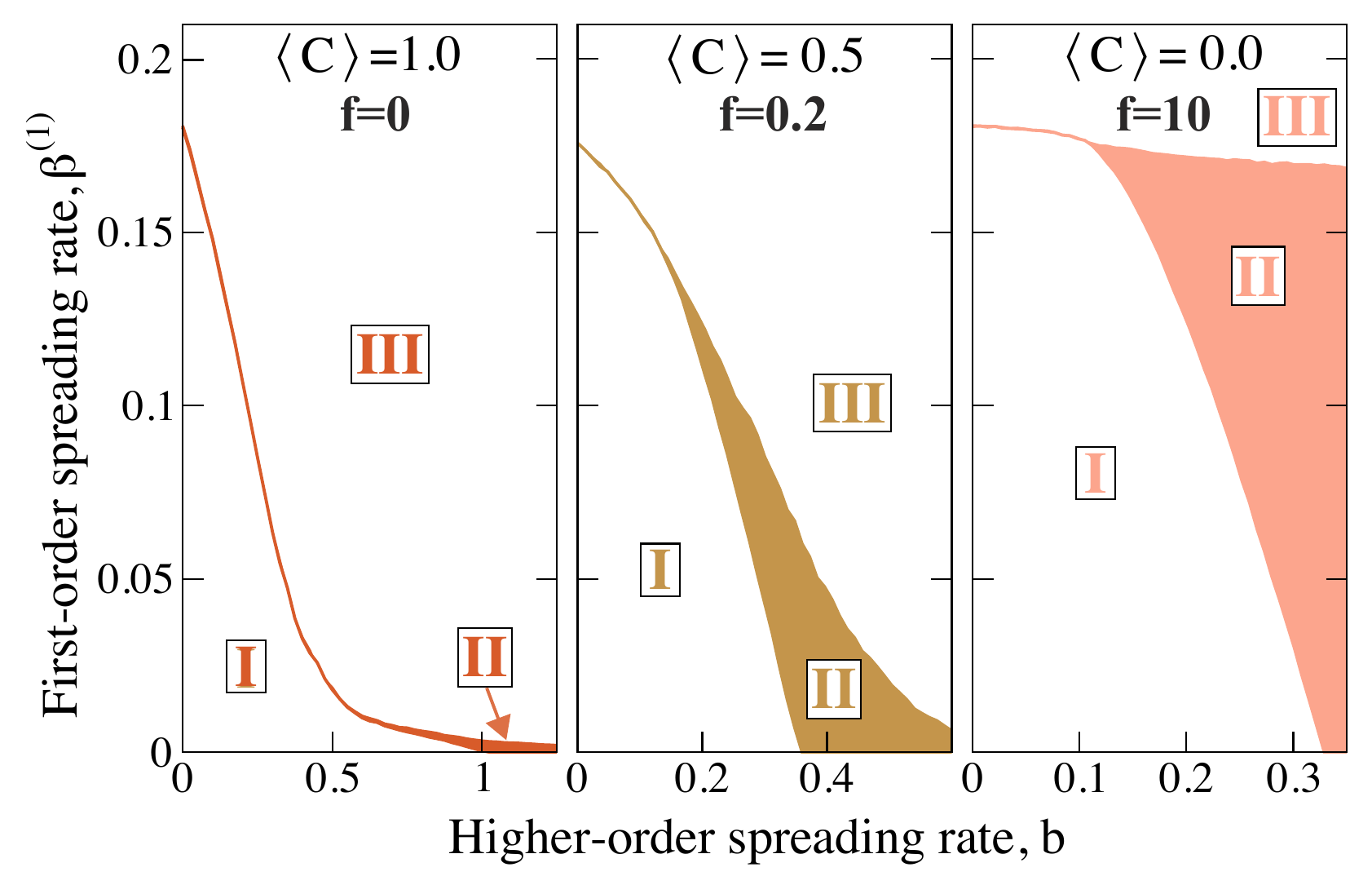}
    \caption{Phase diagrams of the higher-order SIS dynamics in the $(\beta^{(1)}, b)$ parameter space for different levels of embedding (controlled by the rewiring parameter $f$ and corresponding $\langle C \rangle$). Regions denote: I, absorbing phase ($\rho=0$ globally stable); II, bistable regime; III, active phase ($\rho>0$ globally stable). (a) Networks with only triadic interactions. (b) Networks with maximum order $m_{\max}=5$. Both systems have comparable average pairwise degree $\langle k^{(1)} \rangle \approx 8$. Simulations are performed on synthetic networks with $N=10^4$ nodes (see also the SM).}
    \label{fig3}
\end{figure}

\textit{Contagion Dynamics.} We consider a higher-order susceptible–infected–susceptible (SIS) model (hyper-SIS), in which infections can be transmitted by groups of arbitrary size through transitions of the form
\[
m\text{I} + \text{S} \xrightarrow{\beta^{(m)}} (m+1)\text{I},
\]
representing an $m$-order infection process, while recovery occurs via $\text{I} \xrightarrow{\alpha} \text{S}$ as in the standard SIS model~\cite{Barrat2022}. We assume that an $m$-order hyperedge is active when it contains exactly $m$ infected nodes, infecting the remaining susceptible node with rate $\beta^{(m)} = b$. For simplicity, we take a uniform higher-order infection rate $\beta^{(m)} = b$ for all $m>1$, while the pairwise rate $\beta^{(1)}$ is treated as an independent control parameter.

We simulate \cite{maia2025efficient} the dynamics on synthetic hypergraphs with $N=10^4$ nodes, $H=56\,428$ hyperedges, and maximum order $m_{\max}=5$, constructed as described above (see SM). Figure~\ref{fig2} shows the stationary prevalence $\rho$ as a function of $\beta^{(1)}$ for different levels of embedding, with hysteresis probed by initializing the system in low- and high-prevalence states. These results show that embedding strongly reshapes contagion dynamics: highly embedded structures exhibit lower activation thresholds and smooth, continuous transitions, whereas weakly embedded hypergraphs display pronounced hysteresis and discontinuous behavior. Phase diagrams in the $(\beta^{(1)}, b)$ plane (Fig.~\ref{fig3}) further reveal that the bistable region expands as embedding decreases.

While qualitatively consistent with previous studies on limited interaction orders~\cite{Kim2023,Malizia2025,Malizia2026}, our results demonstrate that this behavior extends to heterogeneous hypergraphs with arbitrary order distributions and is governed by a general structural observable. The underlying mechanism is that embedding couples contagion pathways across interaction orders: in highly embedded structures, the activation of higher-order interactions is facilitated by already active lower-order ones, lowering the effective threshold while suppressing the nonlinear amplification required for bistability. In contrast, in weakly embedded hypergraphs, higher-order interactions rely on independent external infections, suppressing activation at low prevalence but enabling abrupt cascades once a critical density is reached, leading to discontinuous transitions. For instance, in a 2-hyperedge, embedding allows pairwise interactions to activate higher-order infection channels already at low prevalence, whereas in weakly embedded structures such activation requires independent external infections. More generally, these mechanistic insights show that discontinuous transitions do not arise in standard pairwise contagion dynamics, but emerge from higher-order interactions that remain sufficiently decoupled, while strong embedding suppresses this effect by coupling activation pathways across orders.

To characterize order-dependent embedding, we define the average nesting coefficient
\begin{equation}
	C^{(M,m)} = \frac{1}{H_M} \sum_{h \,:\, m_h = M} C^{(m)}_{h},
\end{equation}
which quantifies how interactions of order $m$ are embedded within hyperedges of order $M$. This measure captures correlations between interaction orders beyond the average embedding level.

\begin{figure*}[htbp]
	\centering
	\begin{subfigure}[b]{0.58\linewidth}
		\centering
		\includegraphics[width=0.43\textwidth]{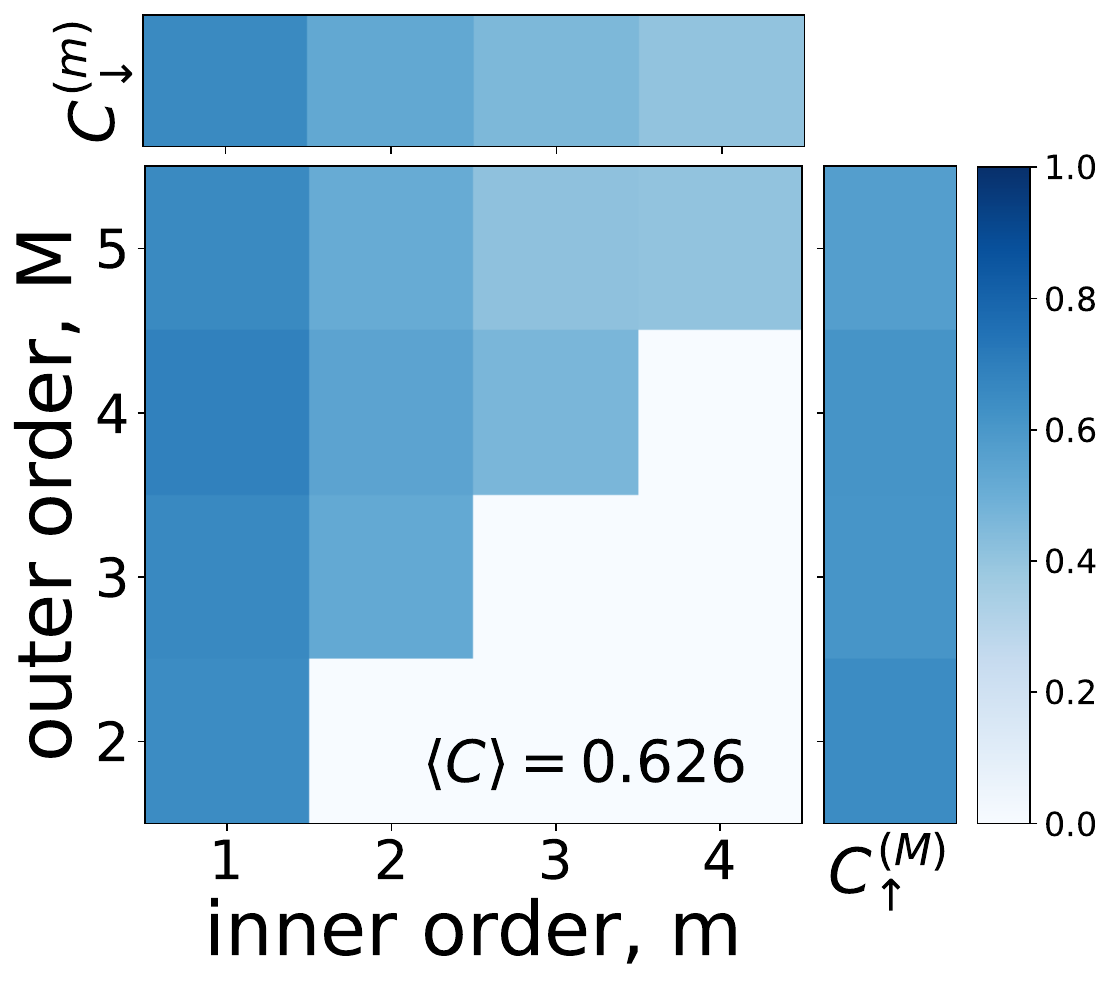} ~~  
		\includegraphics[width=0.43\textwidth]{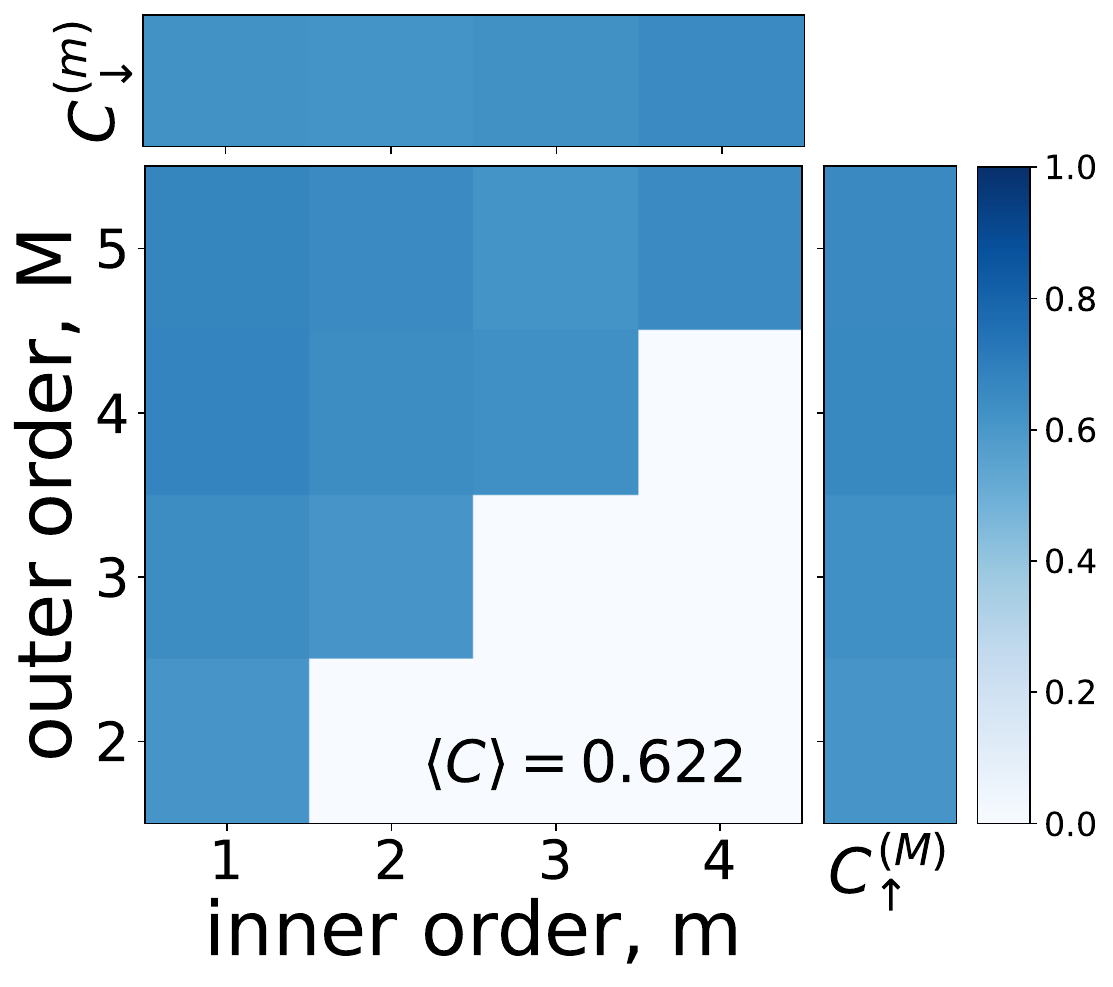} 
		\includegraphics[width=0.43\textwidth]{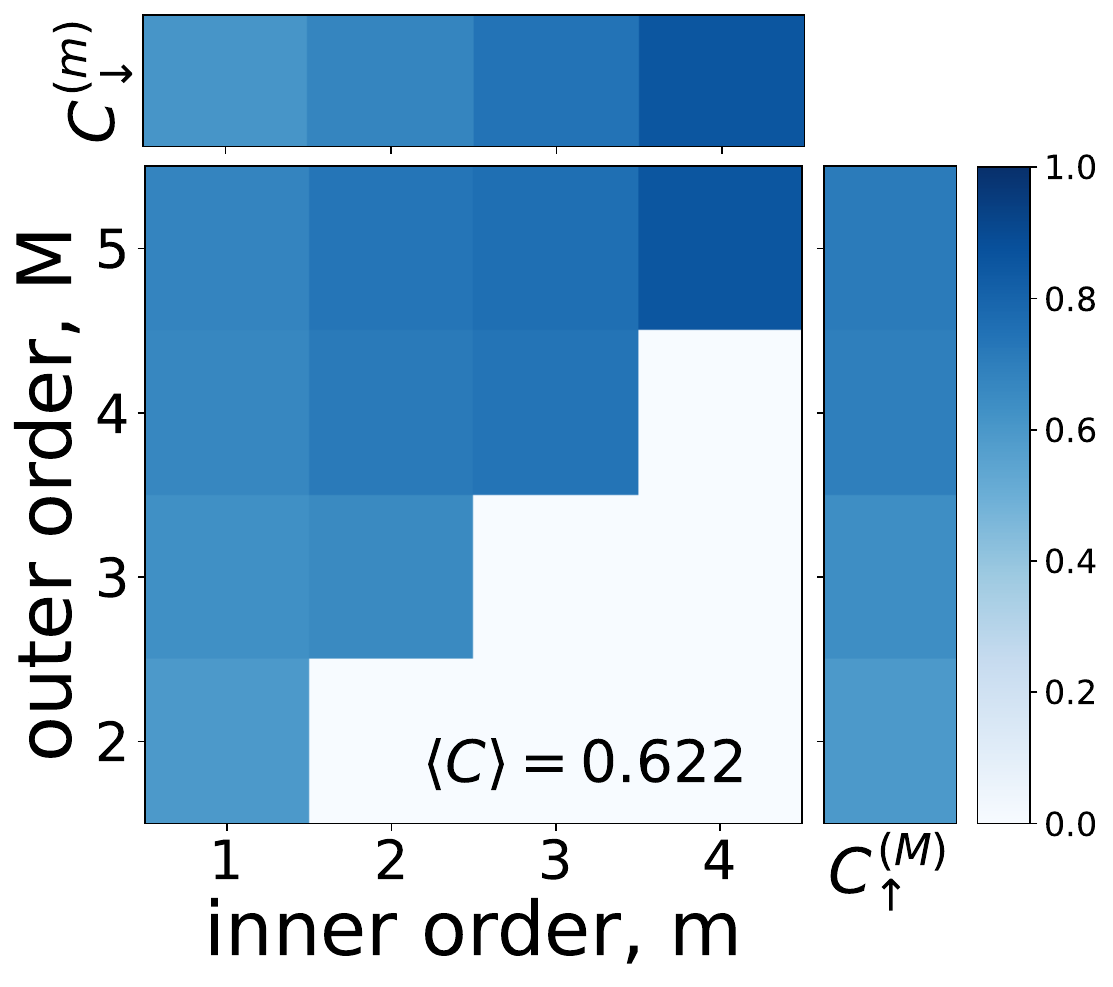}
	\end{subfigure}
	\begin{subfigure}[b]{0.40\linewidth}
		\centering
		\includegraphics[width=\textwidth]{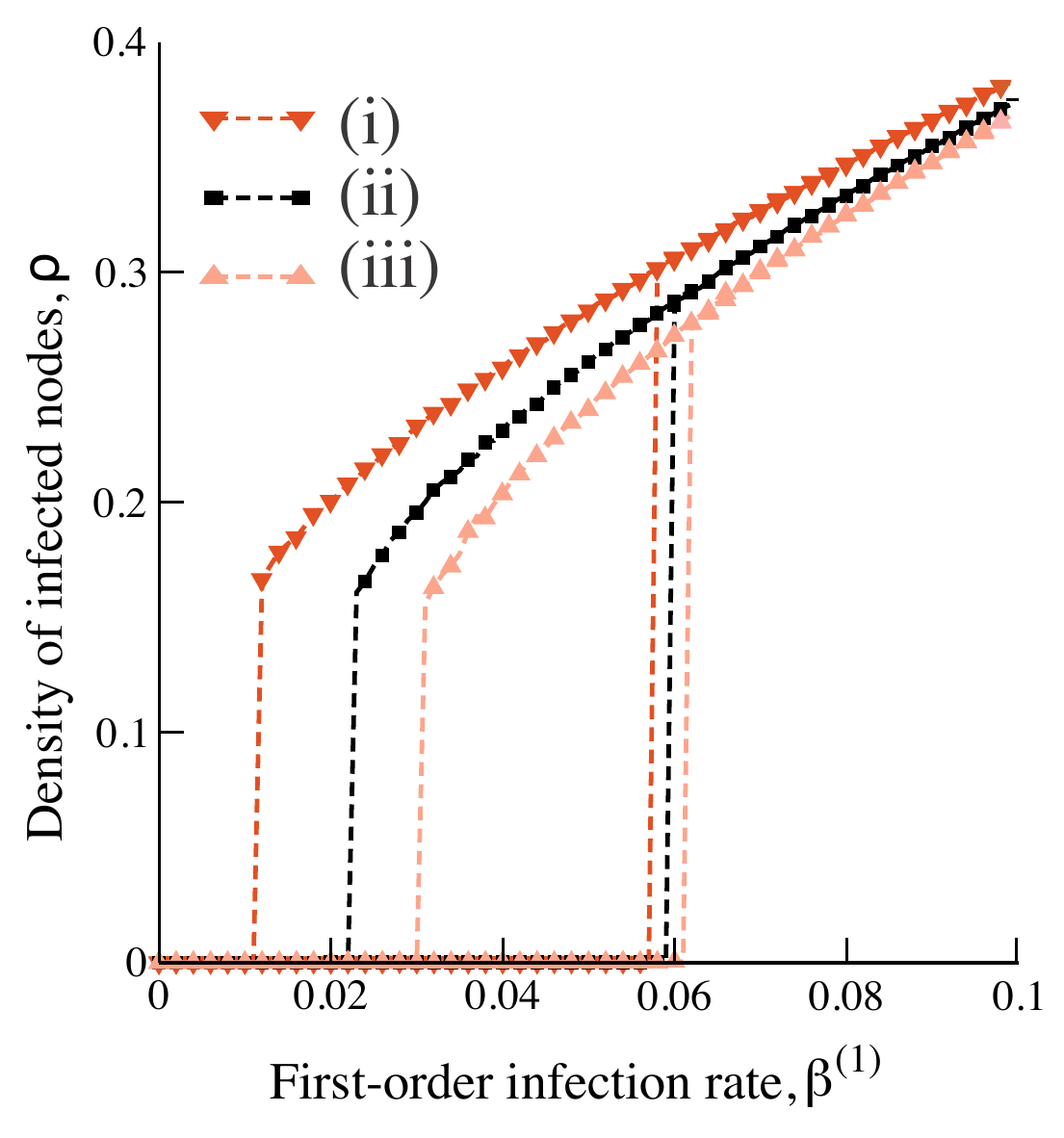}
	\end{subfigure}

    \caption{Effect of order-dependent embedding on higher-order SIS dynamics. (a) Embedding patterns across interaction orders, shown for (i) negative, (ii) neutral, and (iii) positive correlations between embedding and order. Negative correlations correspond to stronger embedding at lower orders, while positive correlations correspond to stronger embedding at higher orders. (b) Stationary prevalence $\rho$ as a function of the first-order spreading rate $\beta^{(1)}$ for the corresponding patterns in (a), showing that embedding concentrated at lower orders facilitates activation while having a limited impact on hysteresis. The higher-order infection rate is $b=0.34$.}
	\label{fig4}
\end{figure*}

We next isolate the impact of these correlations by constructing networks with the same $\langle C \rangle$ but different patterns of $C^{(M,m)}$. Starting from a simplicial complex, we implement three rewiring schemes: random rewiring, preferential rewiring of lower-order hyperedges, and preferential rewiring of higher-order hyperedges. These generate, respectively, approximately uniform embedding, positive correlations between embedding and order, and negative correlations, as shown in Fig.~\ref{fig4}. Negative correlations correspond to stronger embedding at lower interaction orders.

The resulting dynamics reveal that order correlations play a distinct role: negative correlations (i.e., stronger embedding at lower orders) significantly lower the activation threshold, while having only a minor effect on the bistable region and hysteresis width. Conversely, positive correlations hinder activation without substantially enhancing discontinuity. These findings show that contagion dynamics are not determined solely by the overall level of embedding, but also by how embedding is distributed across interaction orders.

\begin{figure*}[htb]
	\begin{subfigure}[b]{0.485\linewidth}
		\centering
		\includegraphics[width=0.485\linewidth]{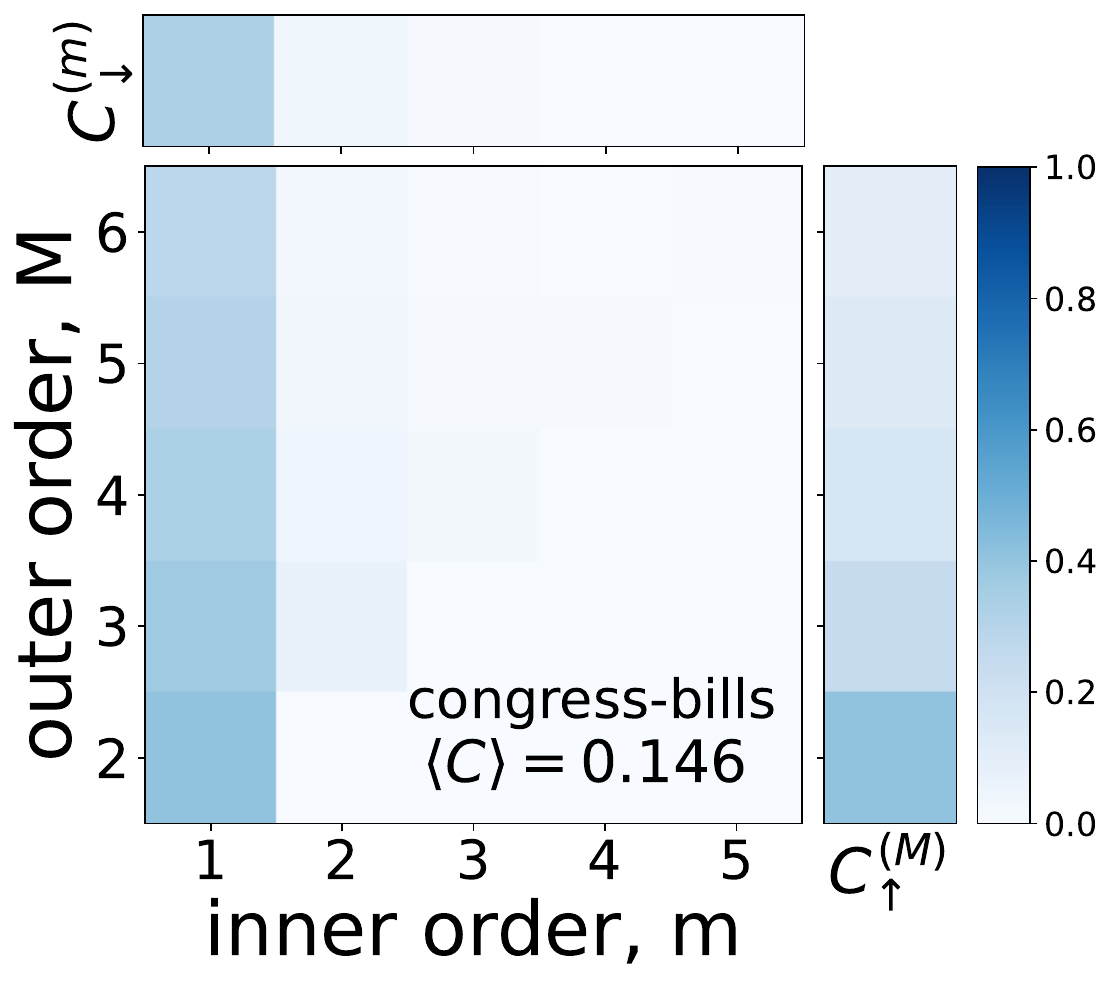}~
		\includegraphics[width=0.485\linewidth]{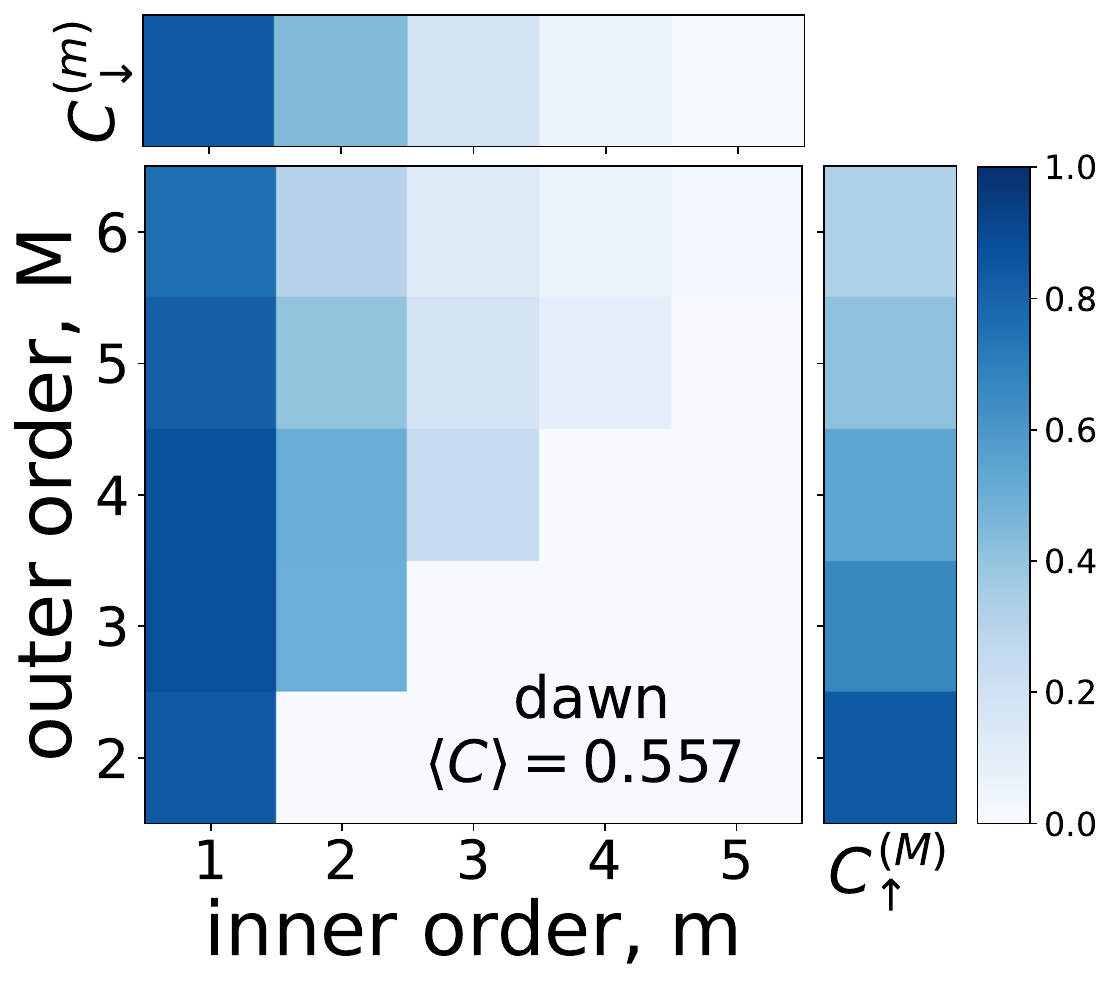}\\
		\includegraphics[width=0.485\linewidth]{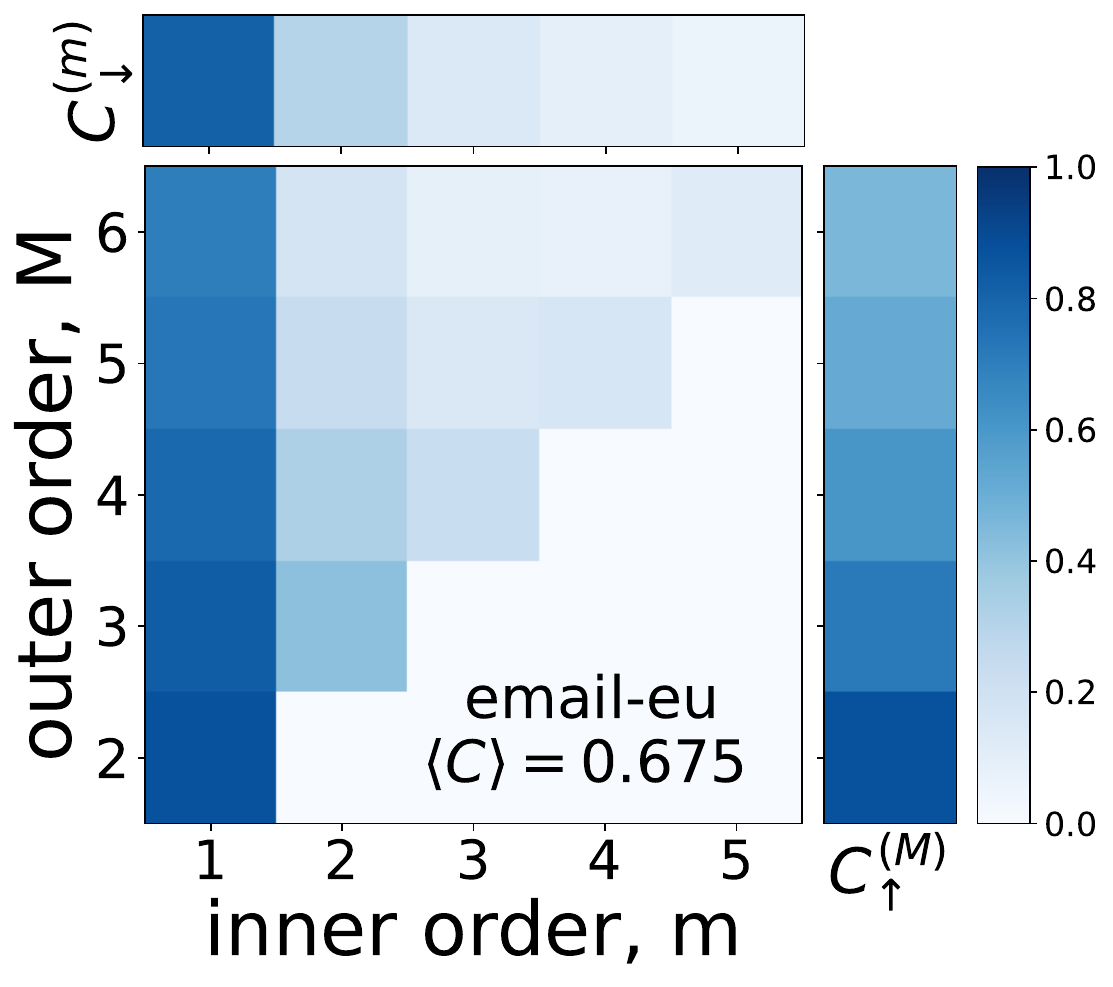}~
		\includegraphics[width=0.485\linewidth]{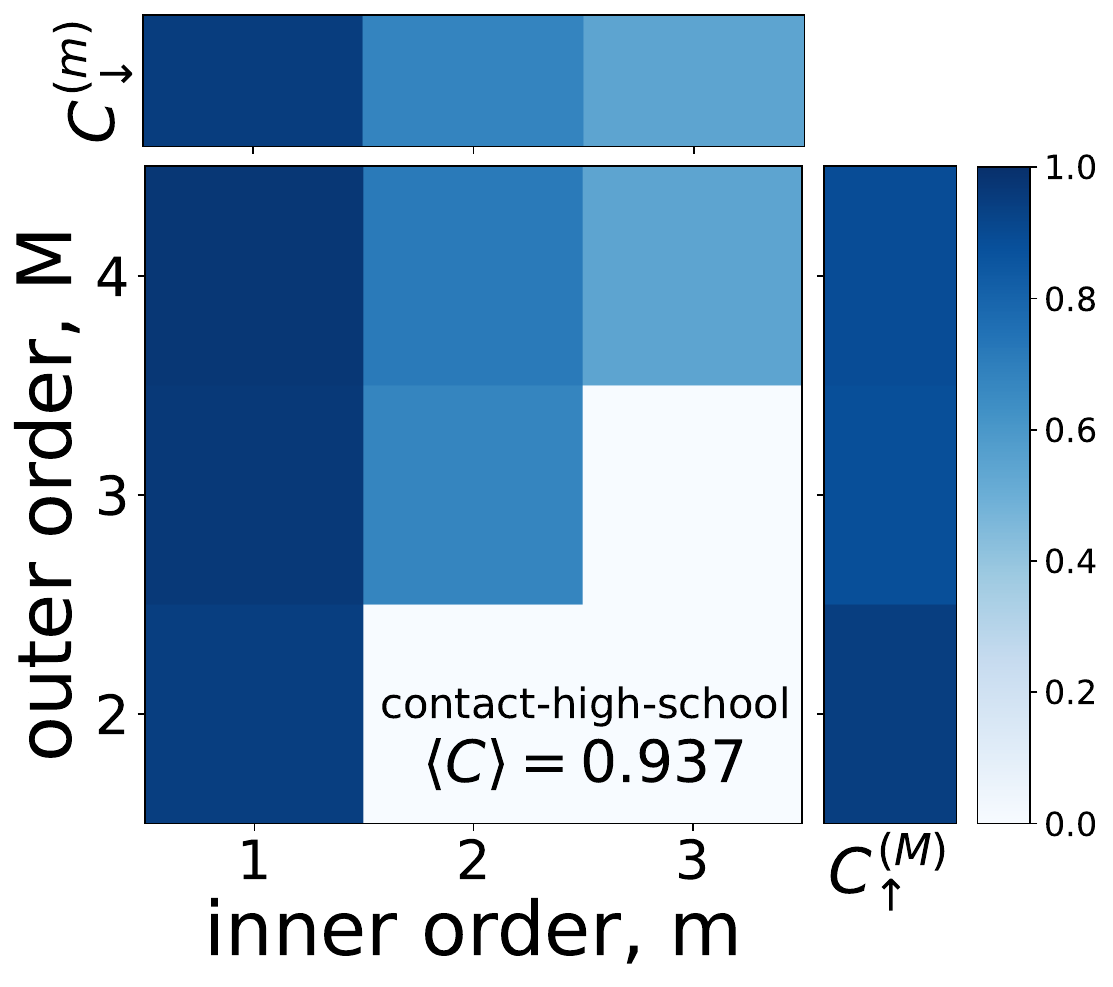}
		\caption{}
		\label{fig5a}
	\end{subfigure}
	\begin{subfigure}[b]{0.485\linewidth}
		\centering
		\includegraphics[width=\linewidth,keepaspectratio]{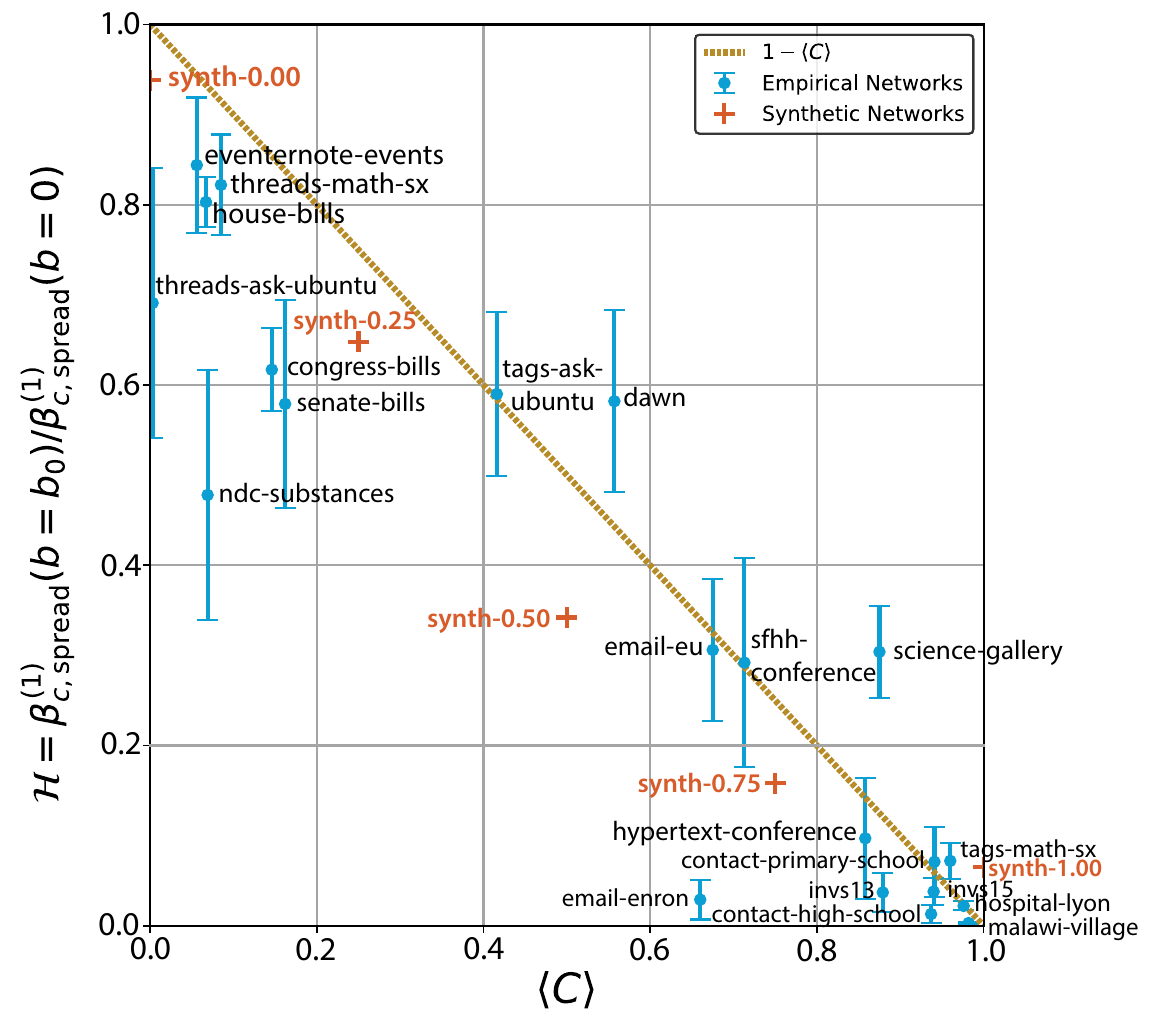}
		\caption{}
		\label{fig5b}
	\end{subfigure}
	
\caption{Empirical evidence of order-dependent embedding and its dynamical impact. (a) Embedding patterns across interaction orders for representative datasets~\cite{landry2023xgi}. Top rows and right columns show the marginal averages over inner- and outer-order embedding, $C^{(m)}_{\downarrow}$ and $C^{(M)}_{\uparrow}$, respectively (see also the SM). The overall embedding $\langle C \rangle$ is indicated in each panel. (b) Scaled bistability range $\mathcal{H} = \beta^{(1)}_{c,\mathrm{spread}}(b=b_0)/\beta^{(1)}_{c,\mathrm{spread}}(b=0)$ as a function of the average embedding $\langle C \rangle$, showing that more embedded networks exhibit reduced hysteresis. Error bars reflect variability in the estimated threshold $\beta^{(1)}_{c,\mathrm{spread}}(b=b_0)$ across stochastic realizations.}
	\label{fig5}
\end{figure*}

\textit{Empirical data --} Having established that both the overall level of embedding and its distribution across interaction orders shape contagion dynamics, we now assess whether these structural features are present in real systems. We analyze a diverse set of higher-order networks, including proximity, online, and biological systems~\cite{stehle2011high, mastrandrea2015contact, vanhems2013estimating, barrat2014measuring, isella2011s, genois2015data, ozella2021using, stehle2011simulation,benson2018simplicial, yin2017local, amburg2020clustering}, obtained from the XGI library~\cite{landry2023xgi}. Figure~\ref{fig5a} shows representative embedding patterns, while the full dataset is reported in the SM. Empirical networks span a broad range of average embedding $\langle C \rangle$ and consistently display negative correlations across interaction orders, indicating stronger embedding at lower orders.

To assess the dynamical implications, we quantify the bistability range through the dimensionless measure $\mathcal{H} = \beta^{(1)}_{c,\mathrm{spread}}(b=b_0)/\beta^{(1)}_{c,\mathrm{spread}}(b=0)$ (see leftmost panel in Fig.~\ref{fig3}a) and compare it with the average embedding. As shown in Fig.~\ref{fig5b}, we find a strong negative correlation between $\mathcal{H}$ and $\langle C \rangle$, indicating that more embedded systems exhibit reduced hysteresis and smoother transitions. This empirical trend directly supports our theoretical predictions.

Finally, to test whether these patterns arise from nontrivial structural organization, we compare each network with a null model that preserves interaction and order distributions while randomizing hyperedge composition. In most cases, the embedding is substantially reduced (see SM), showing that the observed patterns cannot be explained solely by structural constraints but reflect genuine mesoscale organization.

\textit{Conclusions --} We have shown that the organization of higher-order interactions, captured by a nesting coefficient, provides a unifying structural principle governing contagion dynamics on hypergraphs. This measure defines a continuum between simplicial complexes and random hypergraphs and reveals that embedding controls the nature of phase transitions: highly embedded structures lower activation thresholds and promote continuous behavior, whereas weakly embedded systems favor abrupt, discontinuous transitions. Beyond the average level of embedding, we demonstrate that its distribution across interaction orders plays a distinct role: embedding concentrated at lower orders facilitates activation while having a limited impact on hysteresis, highlighting the importance of order-dependent structural correlations. These findings are supported by synthetic and empirical networks, where embedding patterns strongly predict the extent of bistability. More broadly, our results identify embedding as a measurable mesoscale property that bridges structure and dynamics in higher-order systems. By showing that discontinuous transitions arise from sufficiently decoupled higher-order interactions and are suppressed by strong embedding, this work establishes a general mechanism that extends beyond contagion processes and underscores the need for models that explicitly account for higher-order organization.

\paragraph{Acknowledgments.} 

H.P.M and S.C.F  acknowledge the financial support by the  \textit{Fundação de Amparo à Pesquisa do Estado de Minas Gerais} (FAPEMIG)-Brazil (Grants No. APQ-01973-24 and APQ-03079-24) and \textit{Conselho Nacional de Desenvolvimento Científico e Tecnológico} (CNPq)-Brazil (Grants No. 407871/2025-0). 
G.F.A.~ was supported by the \textit{Fundação de Amparo à Pesquisa do Estado de São Paulo} (FAPESP), Process Number 2024/16711-8 and 2025/04409-8. 
S.C.F. is partially supported by CNPq (Grant No. 310984/2023-8), INCT-NeuroComp (CNPq Grant 408389/2024-9), and FAPESP (Grant No. 25/24366-1). 
Y.M. was partially supported by the Government of Arag\'on, Spain, and ``ERDF A way of making Europe'' through grant E36-23R (FENOL), and by Ministerio de Ciencia, Innovaci\'on y Universidades, Agencia Espa\~nola de Investigaci\'on (MICIU/AEI/ 10.13039/501100011033) Grant No. PID2023-149409NB-I00.  
This study was financed in part by the Coordenação de Aperfeiçoamento de Pessoal de Nível Superior - Brasil (CAPES) - Finance Code 001.

%\bibliographystyle{apsrev4-2}
%\bibliography{ref2}


\clearpage
\appendix
%\onecolumngrid
\begin{center}
  \textbf{\large Supplemental Material for "Nesting Controls Phase Transitions in Higher-Order Contagion"}
\end{center}

\section{Additional comments on the Nesting coefficient}

Specific averages of nesting can be measured for the network, an average inner-order nesting ($C^{(m)}_{\downarrow}$) measures how much m-hyperedges are contained in larger hyperedges, an average outer-order nesting ($C^{(M)}_{\uparrow}$) measures how much M-hyperedges contain smaller hyperedges, and an overall average nesting ($\langle C \rangle$) measures total nesting for all orders. These averages are defined, respectively, as,

\begin{equation}
\begin{aligned}
& C^{(m)}_{\downarrow} = \frac{\sum_{M=m+1}^{m_{\text{max}}} H_M C^{(M,m)}}{\sum_{M=m+1}^{m_{\text{max}}} H_M},\\
& C^{(M)}_{\uparrow} = \frac{\sum_{m=1}^{m_{\text{max}}-1} H_m C^{(M,m)}}{\sum_{m=1}^{m_{\text{max}}-1} H_m},\\
& \langle C \rangle = \frac{\sum_{m=1}^{m_{\text{max}}-1} \sum_{M=m+1}^{m_{\text{max}}} H_m H_M C^{(M,m)}}{\sum_{m=1}^{m_{\text{max}}-1} \sum_{M=m+1}^{m_{\text{max}}} H_m H_M}.
\end{aligned}
\end{equation}

Here $H_m$ and $H_M$ are defined as the number of hyperedges of order $m$ and $M$ respectively, while $m_{\text{max}}$ is network's maximum order.

There is in fact an immense difference on the essence of simplicial complexes when compared to hypergraphs, especially when dealing with networks with large maximum order. A single M-order hyperedge comports just one interaction of $M+1$ nodes, while an M-dimensional simplex must necessarily comport $\binom{M+1}{m+1}$ other interactions of a lower-order $m$. This factor is reflected in the network's order distribution, that can manifest in any sort of way for HGs, but a single outlying simplex in a SC with $M_o \gg \langle m \rangle$, for example, would skew the order distribution to a binomial shape with $P_m \sim \binom{M_o+1}{m+1}$. This means that any order distribution that doesn't comport this requisition of existing enough lower-order hyperedges to fill bigger hyperedges will inevitably present nesting closer to zero, making the strict simplicial formalism unrealistic for networks with larger orders.

\section{Generating Synthetic Higher-Order Networks with Varied Nesting}

In this work, synthetic higher-order networks were constructed by means of an adapted version of the Bipartite Confidence Model~\cite{courtney2016generalized} (BCM), which allows for the generation of random networks with predefined interaction and group size distribution. The BCM in itself does not allow for precise control of hyperedge nesting, thus, we opted to generate Simplicial Complexes through the BCM then employ a shuffling mechanism that lowers nesting. In this section we describe the BCM in details and the shuffling mechanism used.

A bipartite graph is formed of two disjointed independent sets of nodes $\{\mathcal{N}^A\}$ and $\{\mathcal{N}^B\}$, each with their own degree distribution $P_k^A$ and $P_k^B$. Nodes from set $A$ are randomly connected to nodes from set $B$, forming two partitions. The resulting bipartite graph is reinterpreted as a higher-order network by considering nodes in partition $A$ as actual nodes or agents in a higher-order network, and nodes in partition $B$ as hyperedges/simplices or groups, with interaction distribution $P_k^A = P_K$ and order/dimension distribution $P_k^B = P_{m+1}$. A link is established from partition $A$ to partition $B$ if the actual node (node in $A$) belongs to the hyperedge/simplex (node in $B$).

\begin{figure}[hbt]
	\begin{center}
		\includegraphics[width=0.95\linewidth]{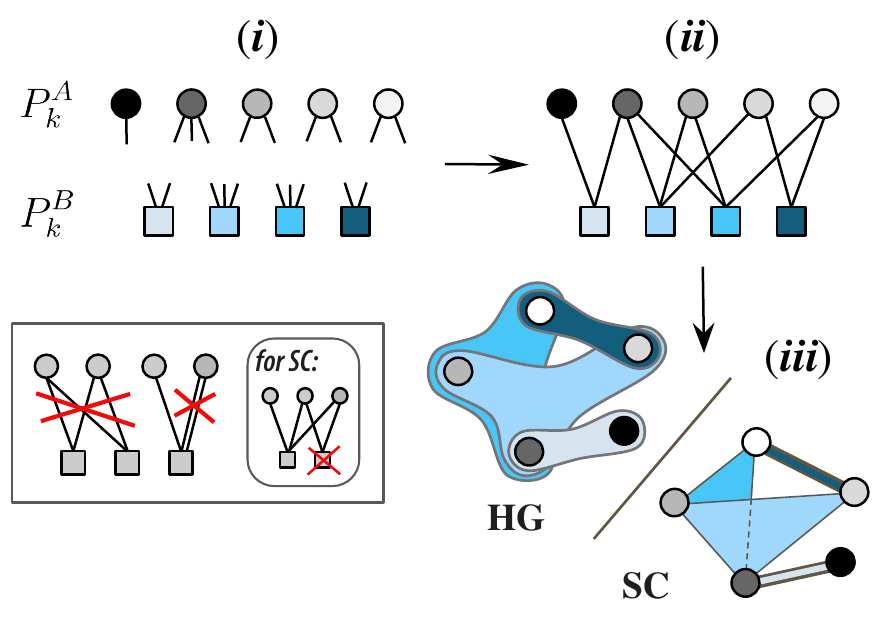}
	\end{center}
	\caption{ Example of the three steps employed to generate networks through the bipartite configuration model (see SM text). In this example, a bipartite graph is built with five nodes in partition $A$ (circles) and four nodes in partition $B$ (squares) following from predefined degree distributions $P_k^A$ and $P_k^B$. Nodes from each partition can connect only once to each other, and nodes from partition $B$ with the same degree cannot possess the same set of neighbors. The resulting graph is reinterpreted as a higher-order network with five nodes and four hyperedges.}
	\label{fig:BipartiteCM}
\end{figure}

The BCM follows three steps to generate higher-order networks. The steps are listed below and illustrated in figure~\ref{fig:BipartiteCM}. Step \textbf{(i)}, create two sets of nodes with predefined distributions $P_k^A$ and $P_k^B$. The total sum of degrees in $A$ must be equal to that of $B$, $S_k^A = \sum_{i \in A}^{N_A} k_i = S_k^B = \sum_{i \in B}^{N_B} k_i$. One recommended approach is to first define $N_A$ and distribute their degrees following $P_k^A$, thus obtaining $S_k^A$. Then nodes are subsequently added to $B$ with degrees drawn from $P_k^B$ until the condition $S_k^A = S_k^B$ is met. Step \textbf{(ii)}, randomly connect nodes from both partitions until their number of connections matches their degrees. Step \textbf{(iii)}, the resulting bipartite graph is reinterpreted as a higher-order network with predefined interaction and order distributions.

Since a hyperedge cannot contain the same node multiple times, nodes from partition $A$ and $B$ must connect to each other at most only once. Due to this constraint, randomly wiring nodes from both partitions might result in frustrated configurations with nodes that can't complete their connections. One last detail to consider is that two or more hyperedges of the same order might contain the same sets of nodes. These are instances of different groups formed by the exact same agents, and it can be sidestepped by selecting a repeated hyperedges and swapping at least one of its nodes with another random hyperedge. This of course provided that it doesn't violate any previous rule.

For the creation of SC, we start with the construction of a hypergraph with nodes and hyperedges connected according to the predefined distributions. Then, all subfaces of smaller-orders within hypergraphs are added to the bipartite representation, turning all higher-order interactions into simplices. It is important to note that the degree and order distributions will be altered by this transformation into SC, reshaping the distribution into something closer to a binomial distribution.

To isolate the effects of nesting in contagion dynamics, we employ a rewiring mechanism that when applied to SC, creates higher-order networks with reduced nesting.

A rewire is executed by simply selecting a pair of hyperedges of the same order then swapping random nodes between them. The restriction of selecting hyperedges with equal order is done with the intention to maintain the precise m-degree distribution of individual nodes. To parametrize the amount of hyperedges rewired, a total of $f \times NH$ pairs of hyperedges have their nodes swapped, with $f$ being any number greater than zero, defined as a fraction of hyperedge rewires. This rewiring mechanism retains the original network's interaction, order and specific $m$-degree distribution.

\begin{figure}[hbt]
	\begin{center}
		\includegraphics[width=0.9\linewidth]{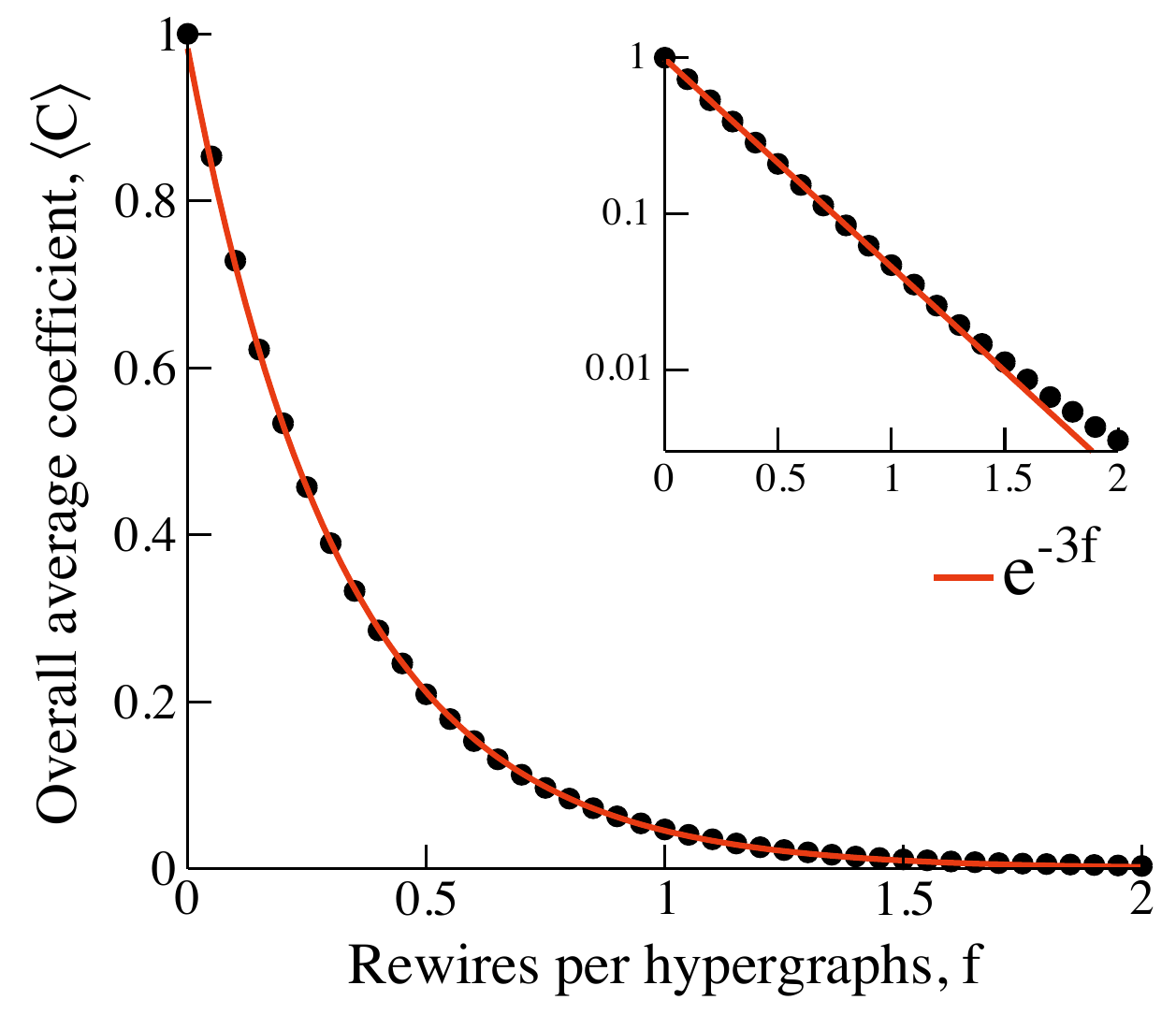}
	\end{center}
	\caption{ Overall average nesting coefficient $\av{C}$ as a function of rewires per hypergraphs $f$. Starting from a SC with $\av{C} = 1$, the nesting coefficient drops exponentially as a function of the amount of rewires per hyperedge. }
	\label{fig:Caverage}
\end{figure}

As shown in figure~\ref{fig:Caverage}, increasing $f$ reduces $\av{C}$ exponentially, with the decay exponent being determined by the original network's structure. With high enough values of $f$, the network has no significant $\av{C}$, therefore being in a regime of nesting coefficients similar to random HG.

In the main text, two networks were created for figure 3. For figure 3a, a regular Simplicial Complex is created with each node connected to 4 2-simplices, resulting in a pairwise degree of $\langle k^{(m=1)} \rangle = 8$, a triad degree of $\langle k^{(m=2)} \rangle = 4$ and an average order of $\langle m \rangle = 1.25$. For figure 3b, a Simplicial Complex is created with all nodes being connected to 3 simplices at random. Simplices are distributed from orders 1 to 5 with proportions 52\% of order 1, 29\% of order 2, 13\% of order 3, 5\% of order 4 and 1\% of order 5. This results in an average pairwise degree of $\langle k^{(m=1)} \rangle \approx 8$ and an average order of $\langle m \rangle = 1.72$.

For the creation of the networks used in figure 4 of the main text, we introduce a bias in selecting the orders of the hyperedges that will be rewired. The network structure is the same as the figure 3b. For a bias of more nesting in lower orders, rewiring bigger orders should be favored, so a randomly chosen hyperedge should be rewired with probability proportional to $m/m_{\text{max}}$. For a bias of more nesting in higher-orders, rewiring smaller hyperedges should be favored, therefore rewiring should be in turn proportional to $1 - m/m_{\text{max}}$. For balanced values of nesting, no bias is induced and hyperedges are selected to be rewired with equal chance.

\section{Full Empirical Data}

In the main text, the nesting coefficients was shown for only a few selected datasets. Here we include a full figure with all the measured nesting coefficients. Additionally, a table with numerical values for the average nesting coefficients and other parameters including the ones used to measure the hysteresis length $\mathcal{H}$ are also shown.

\begin{figure*}[htbp]
    \centering
    \begin{subfigure}[b]{0.24\textwidth}
        \centering
        \includegraphics[width=\linewidth]{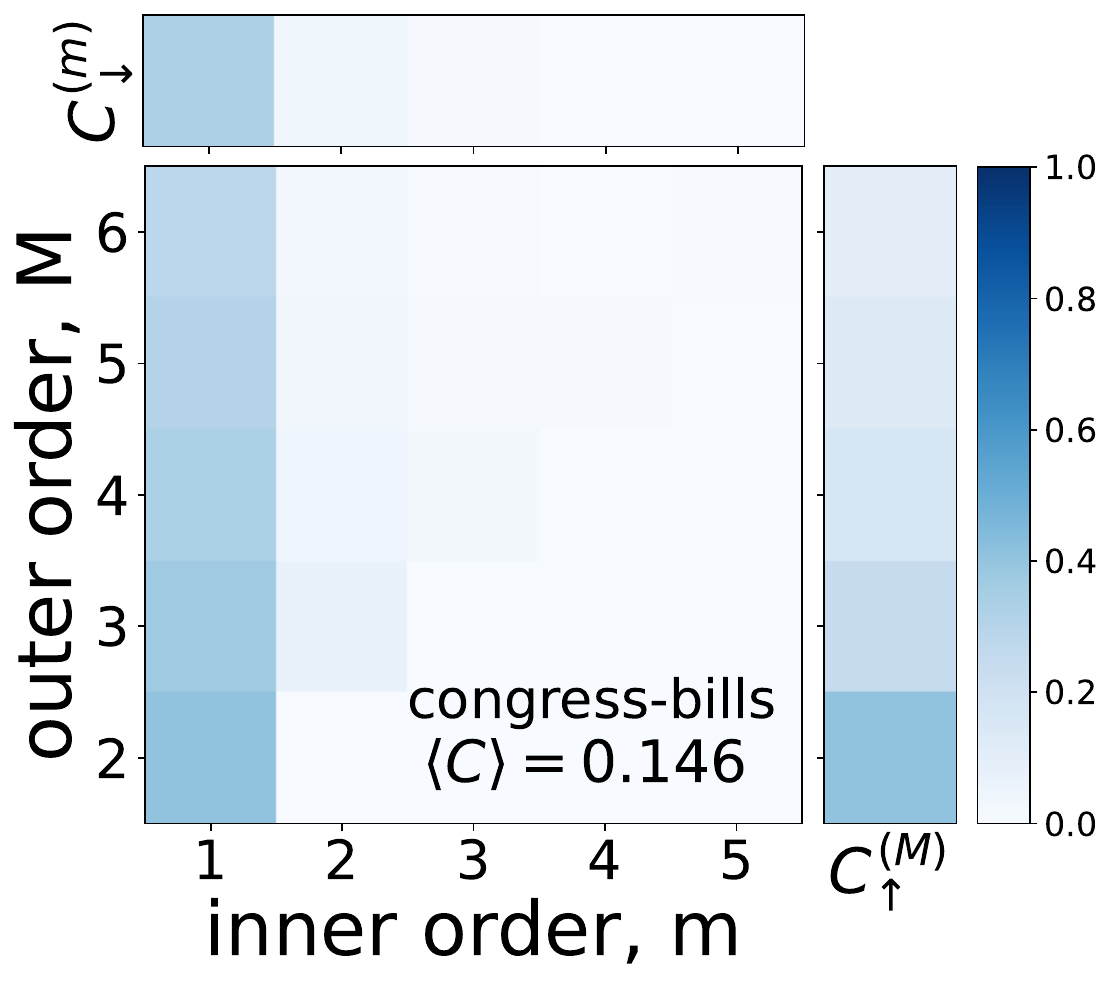}
    \end{subfigure}
    \hfill
    \begin{subfigure}[b]{0.24\textwidth}
        \centering
        \includegraphics[width=\linewidth]{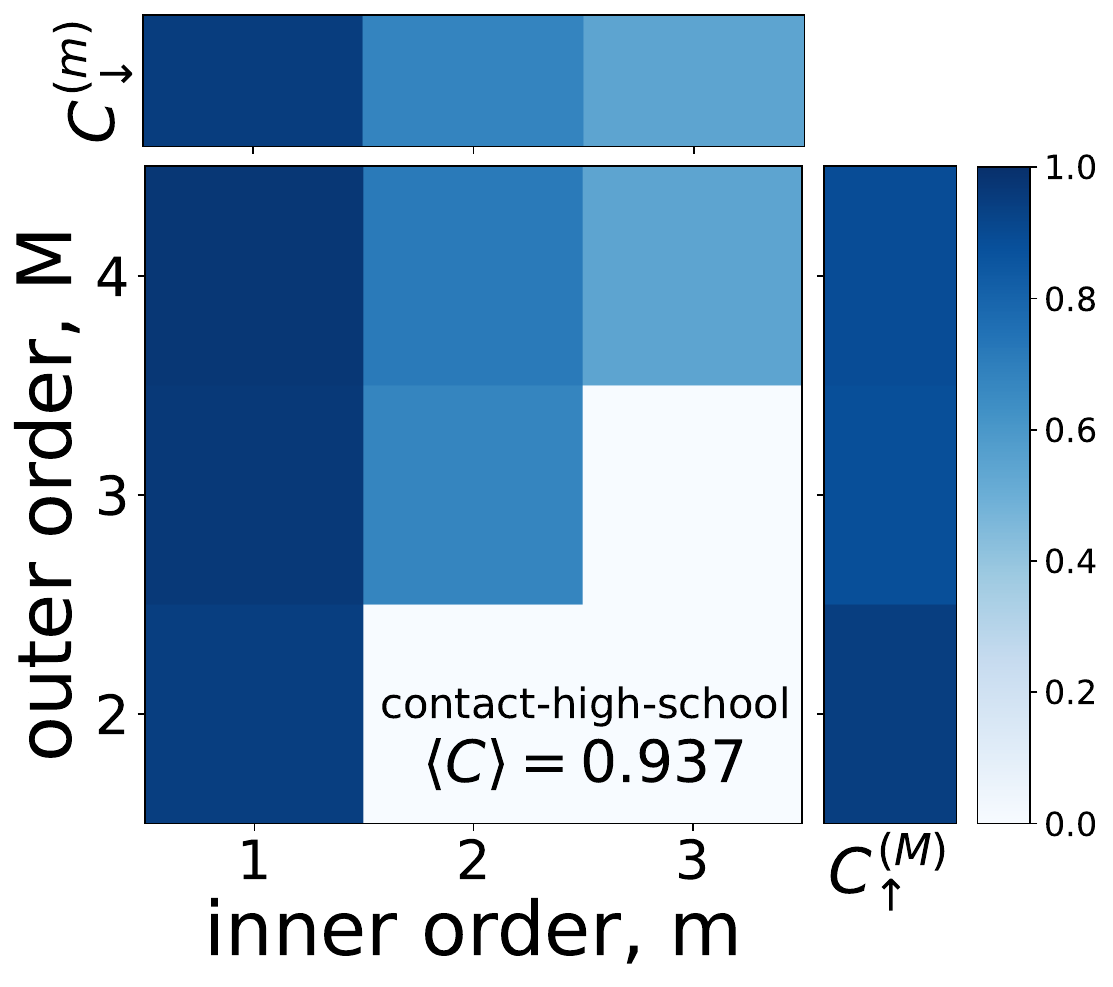}
    \end{subfigure}
    \hfill
    \begin{subfigure}[b]{0.24\textwidth}
        \centering
        \includegraphics[width=\linewidth]{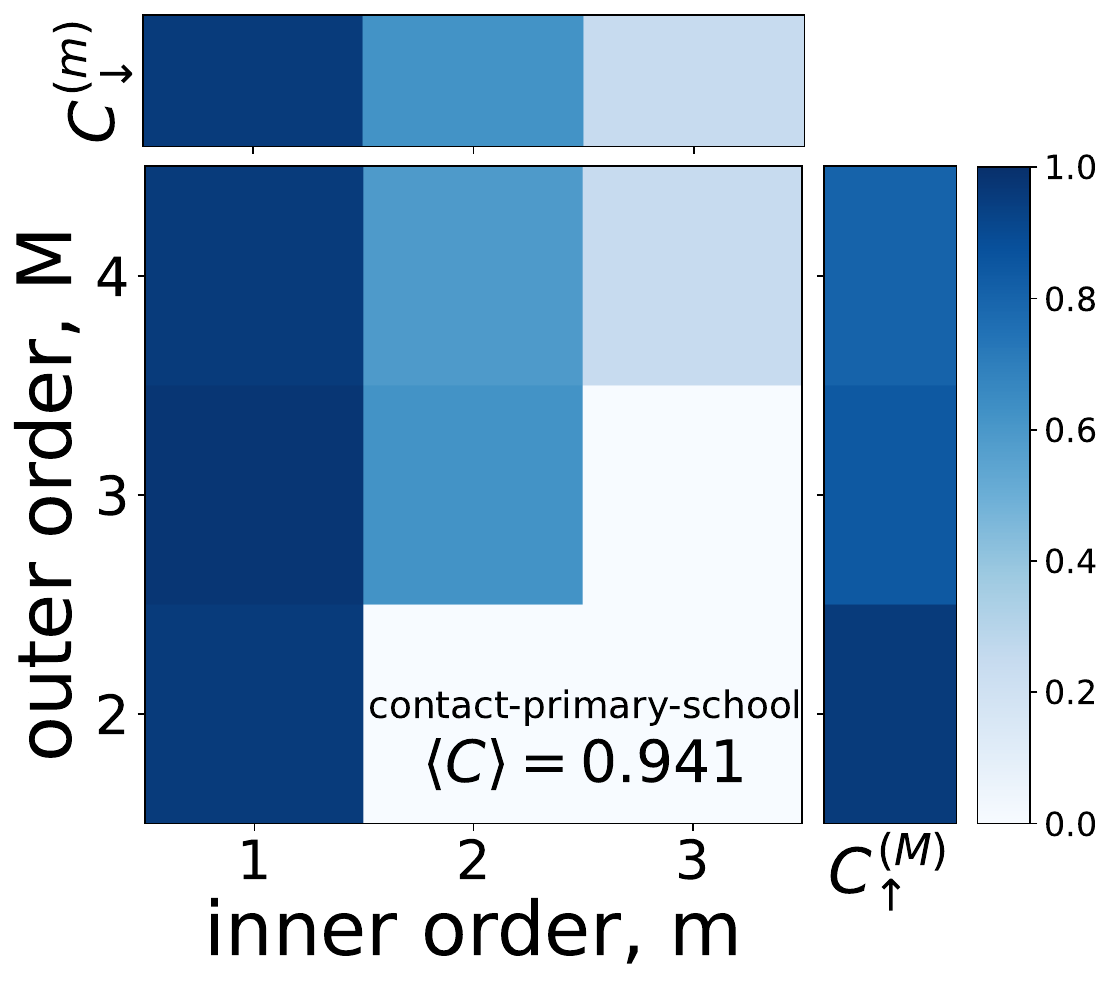}
    \end{subfigure}
    \hfill
    \begin{subfigure}[b]{0.24\textwidth}
        \centering
        \includegraphics[width=\linewidth]{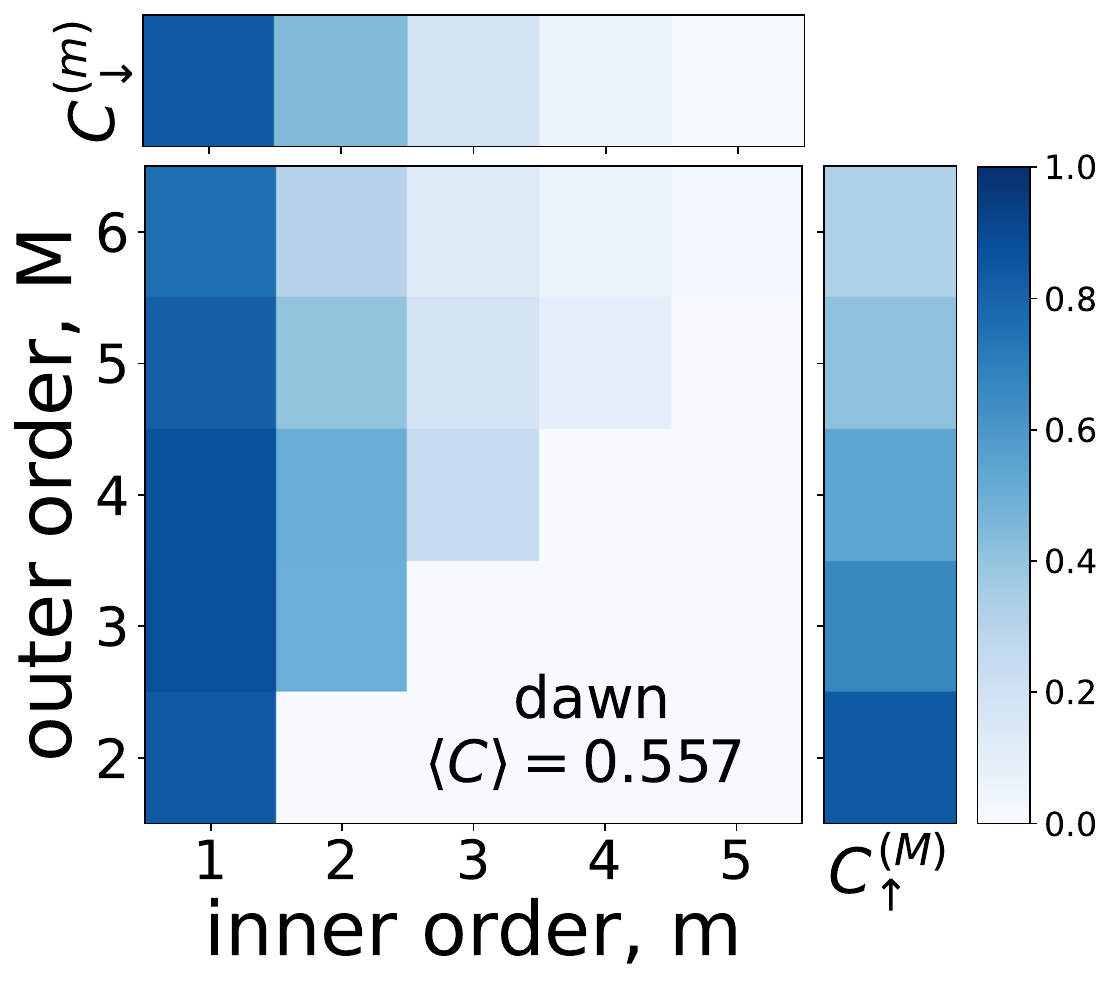}
    \end{subfigure}
    \hfill
    \begin{subfigure}[b]{0.24\textwidth}
        \centering
        \includegraphics[width=\linewidth]{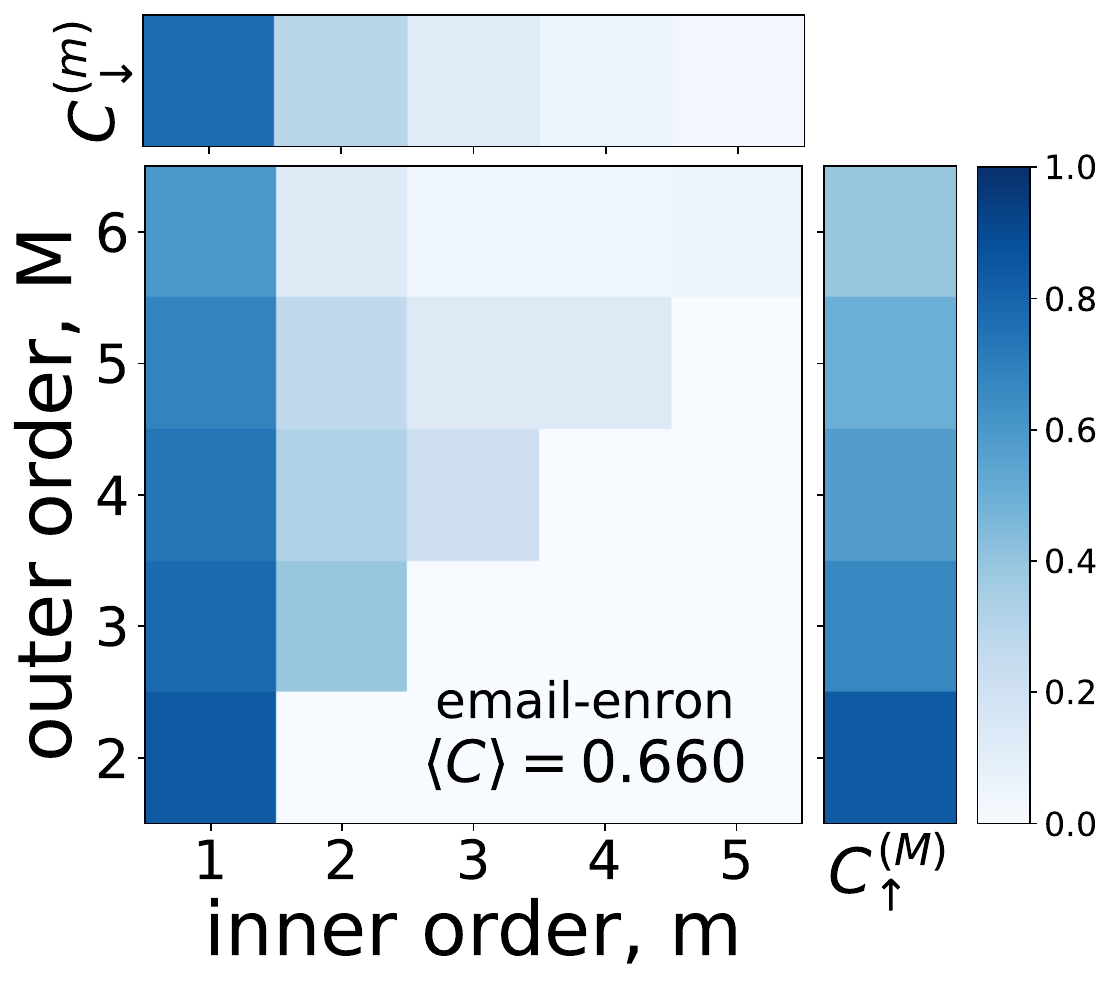}
    \end{subfigure}
    \hfill
    \begin{subfigure}[b]{0.24\textwidth}
        \centering
        \includegraphics[width=\linewidth]{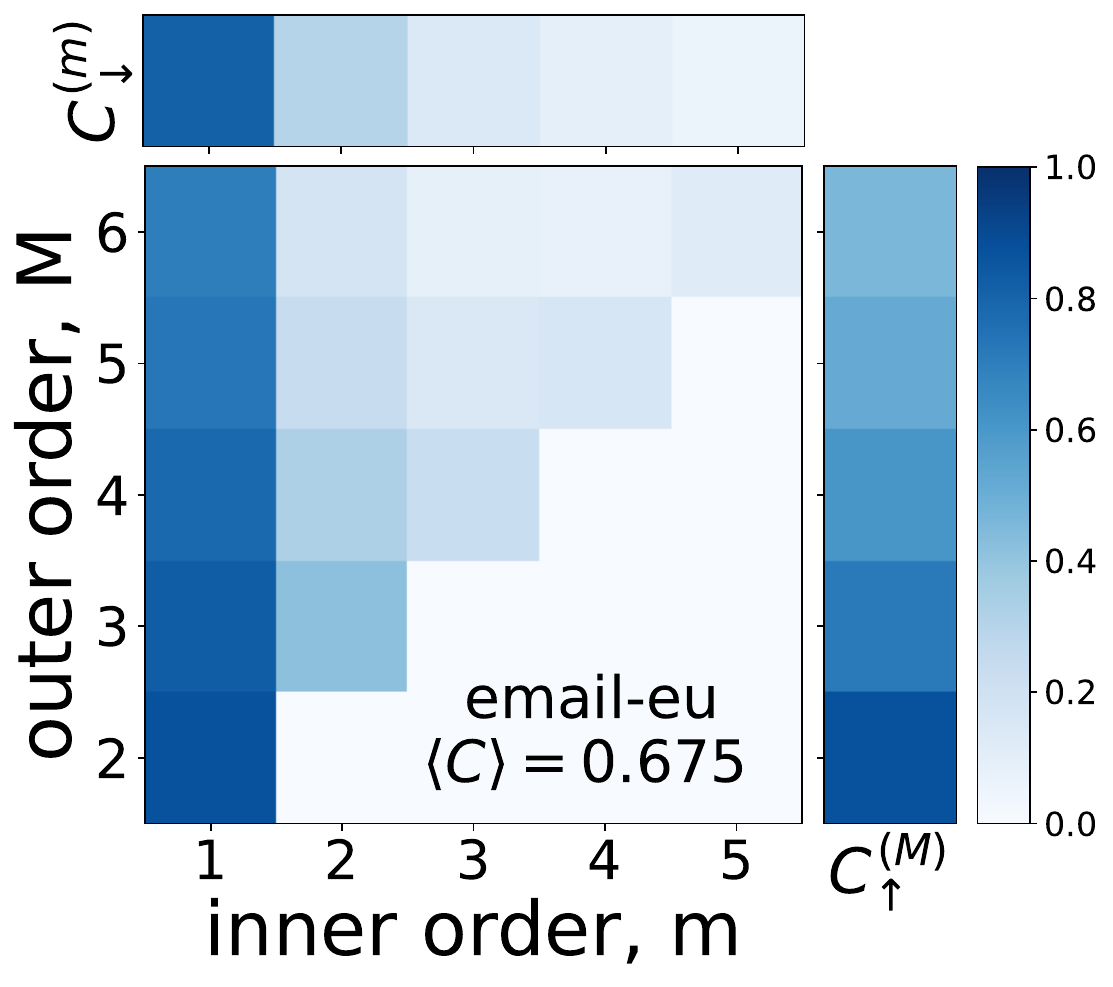}
    \end{subfigure}
    \hfill
    \begin{subfigure}[b]{0.24\textwidth}
        \centering
        \includegraphics[width=\linewidth]{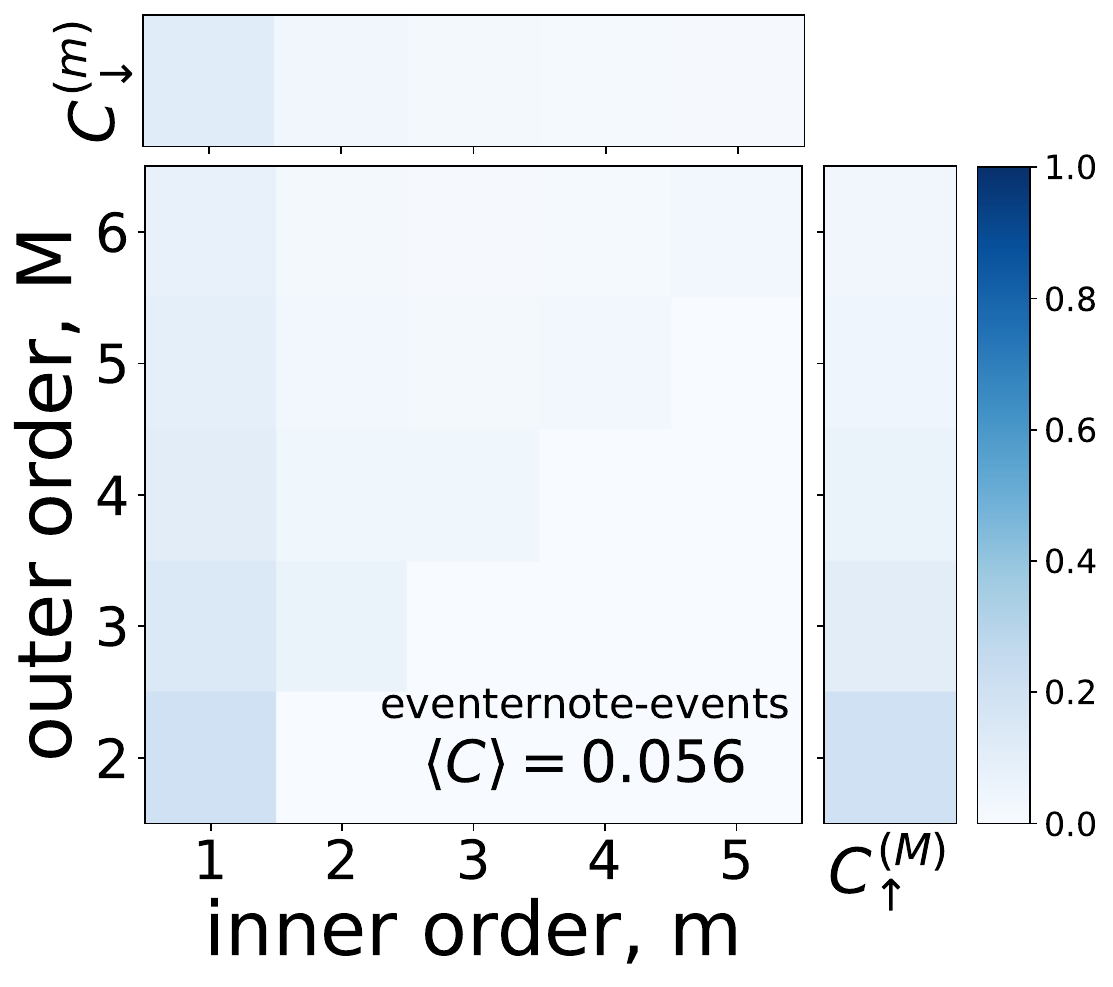}
    \end{subfigure}
    \hfill
    \begin{subfigure}[b]{0.24\textwidth}
        \centering
        \includegraphics[width=\linewidth]{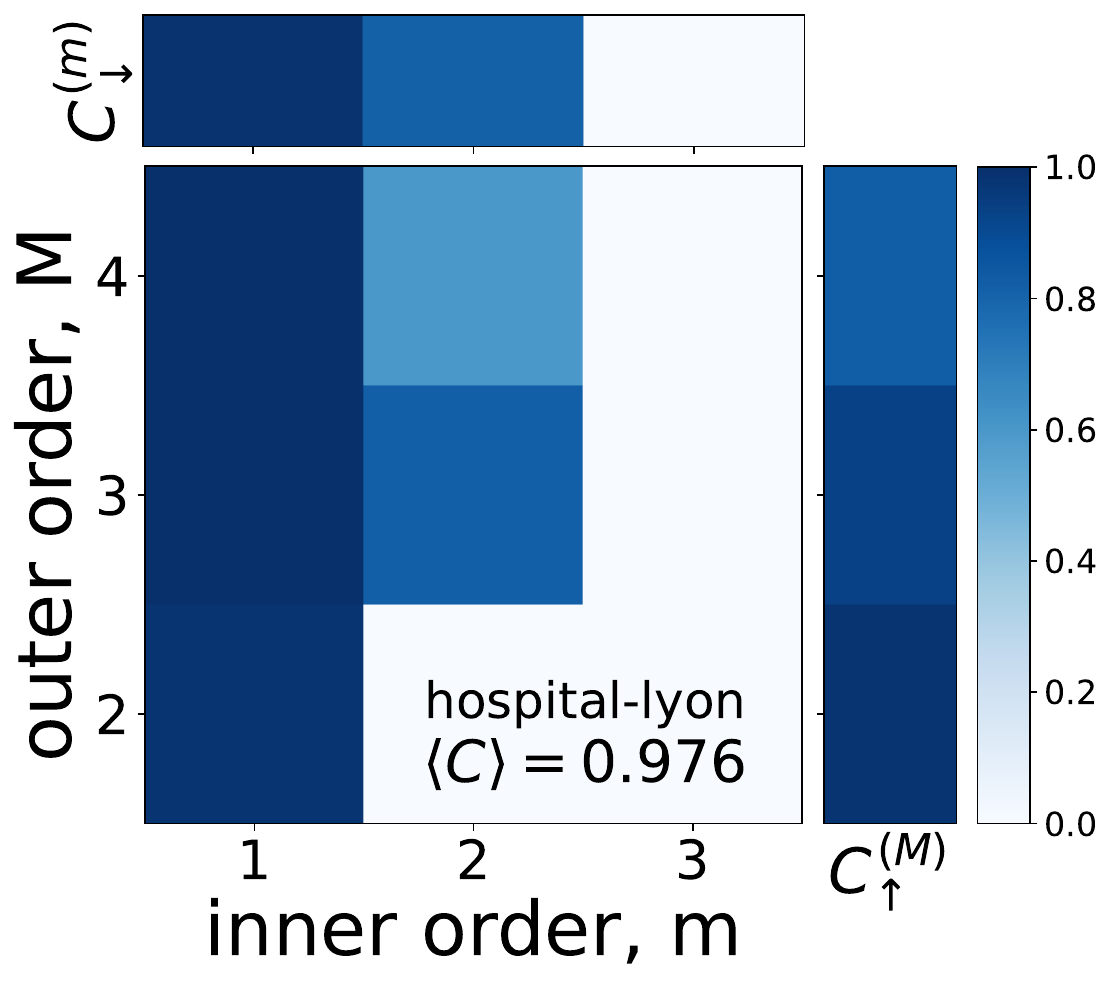}
    \end{subfigure}
    \hfill
    \begin{subfigure}[b]{0.24\textwidth}
        \centering
        \includegraphics[width=\linewidth]{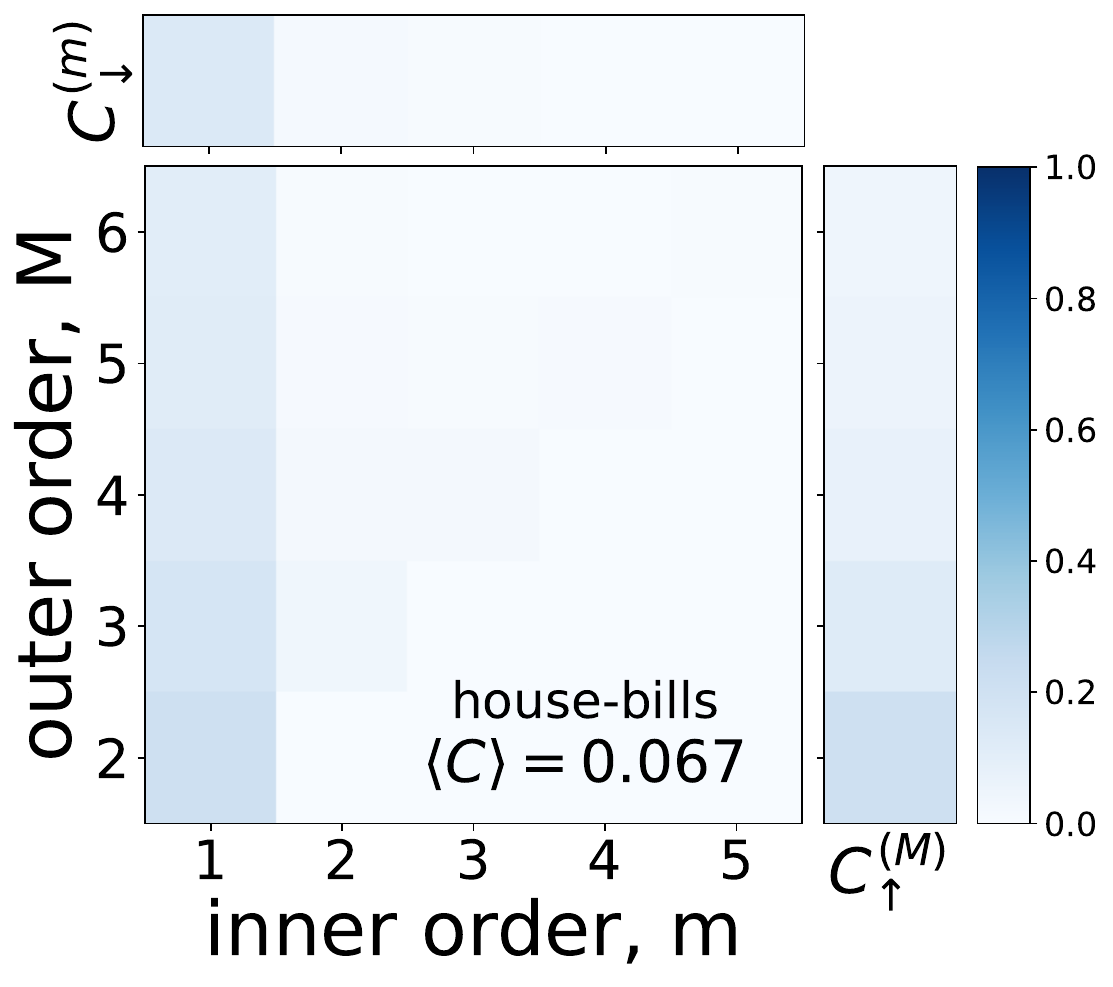}
    \end{subfigure}
    \hfill
    \begin{subfigure}[b]{0.24\textwidth}
        \centering
        \includegraphics[width=\linewidth]{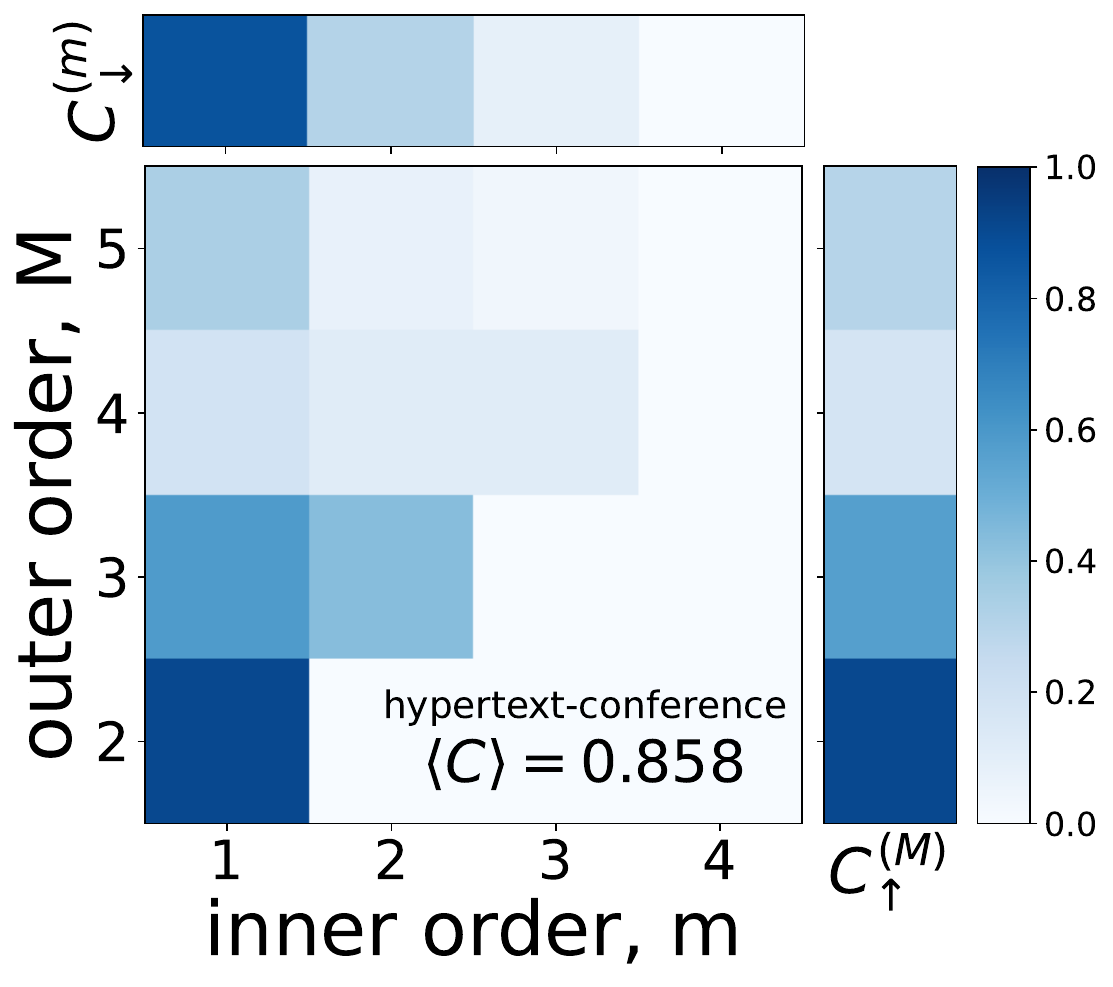}
    \end{subfigure}
    \hfill
    \begin{subfigure}[b]{0.24\textwidth}
        \centering
        \includegraphics[width=\linewidth]{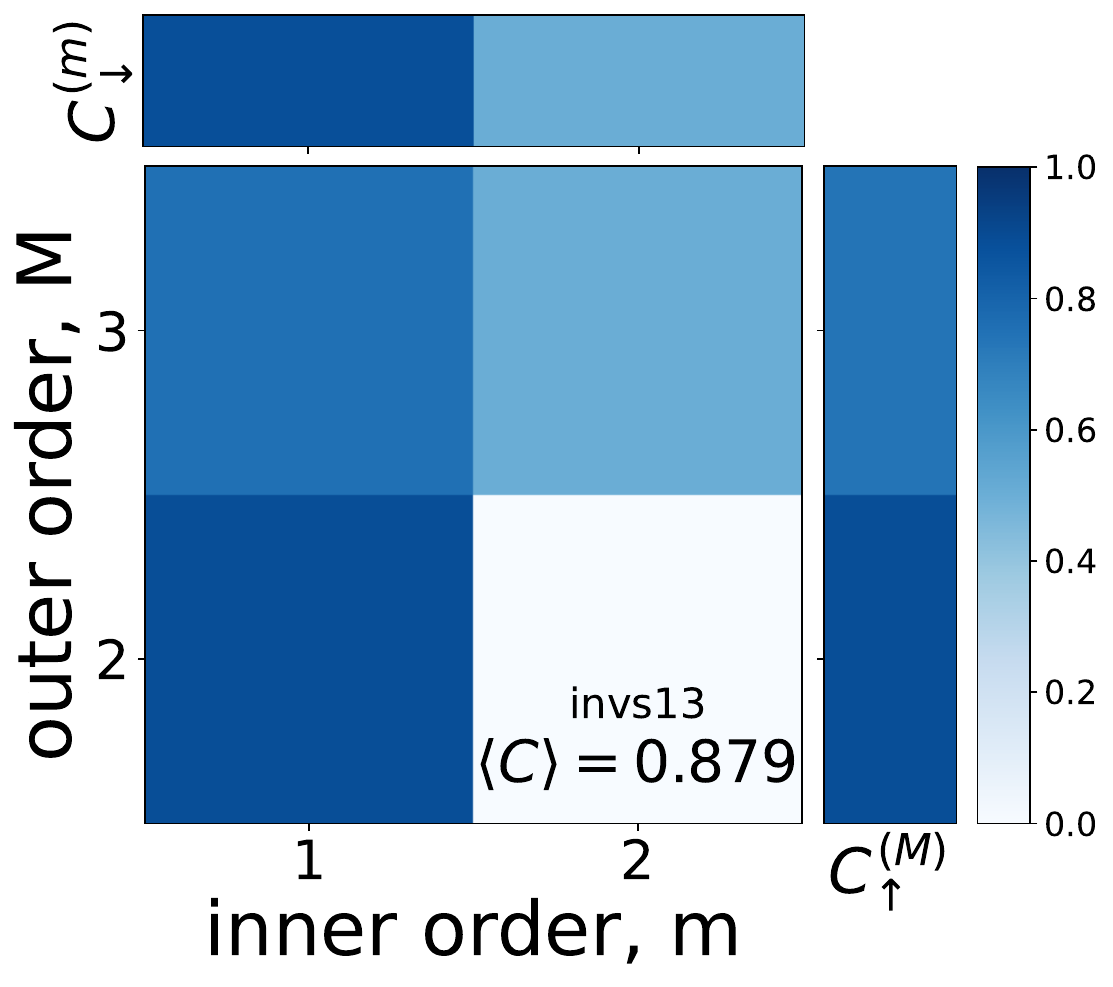}
    \end{subfigure}
    \hfill
    \begin{subfigure}[b]{0.24\textwidth}
        \centering
        \includegraphics[width=\linewidth]{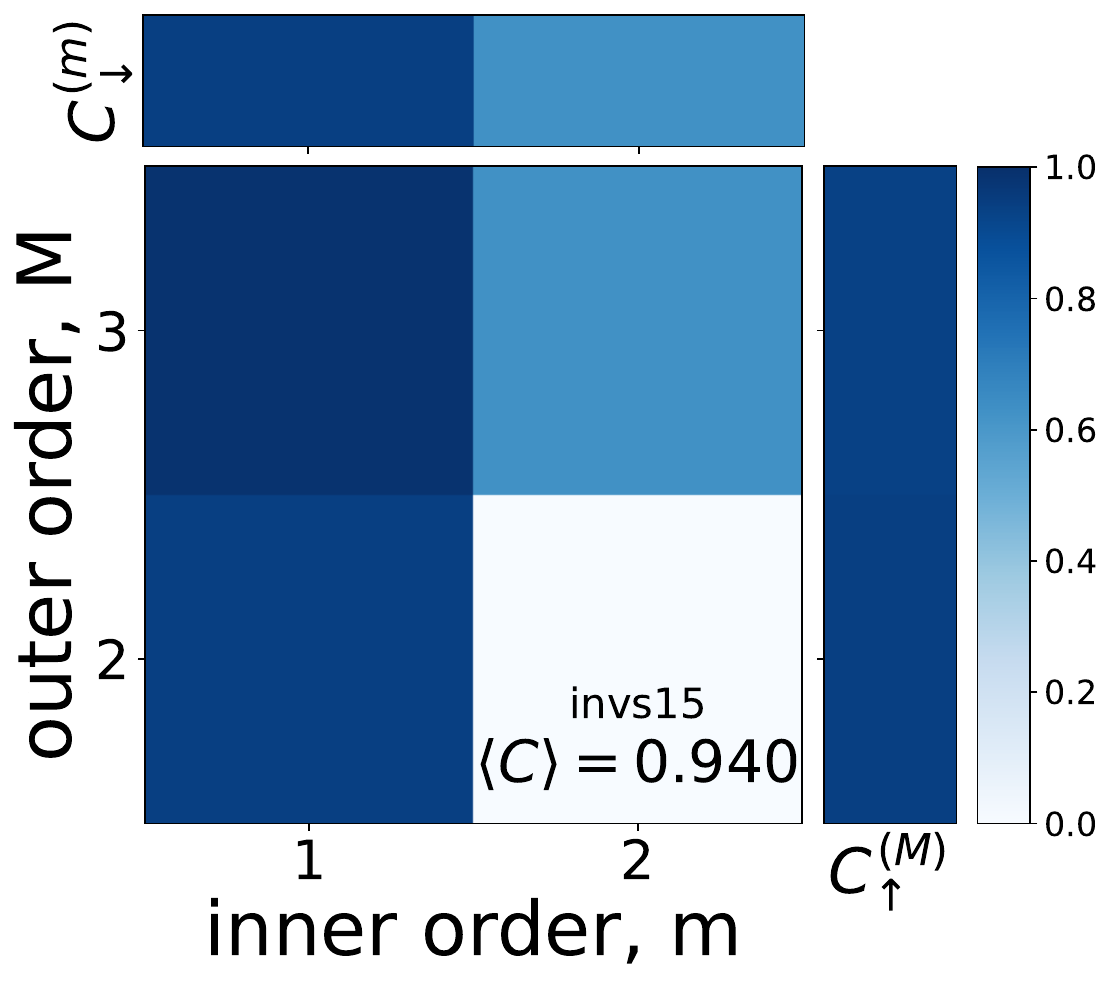}
    \end{subfigure}
    \hfill
    \begin{subfigure}[b]{0.24\textwidth}
        \centering
        \includegraphics[width=\linewidth]{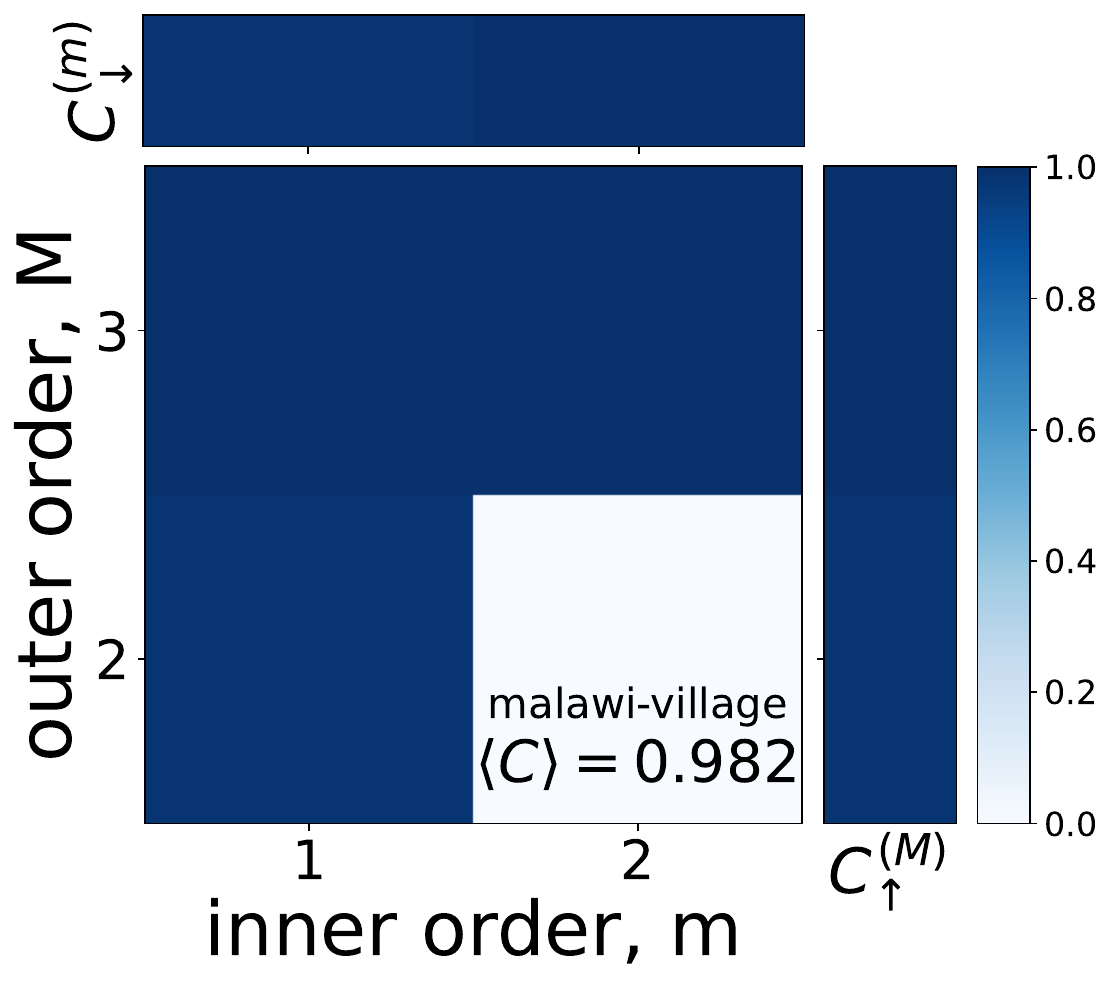}
    \end{subfigure}
    \hfill
    \begin{subfigure}[b]{0.24\textwidth}
        \centering
        \includegraphics[width=\linewidth]{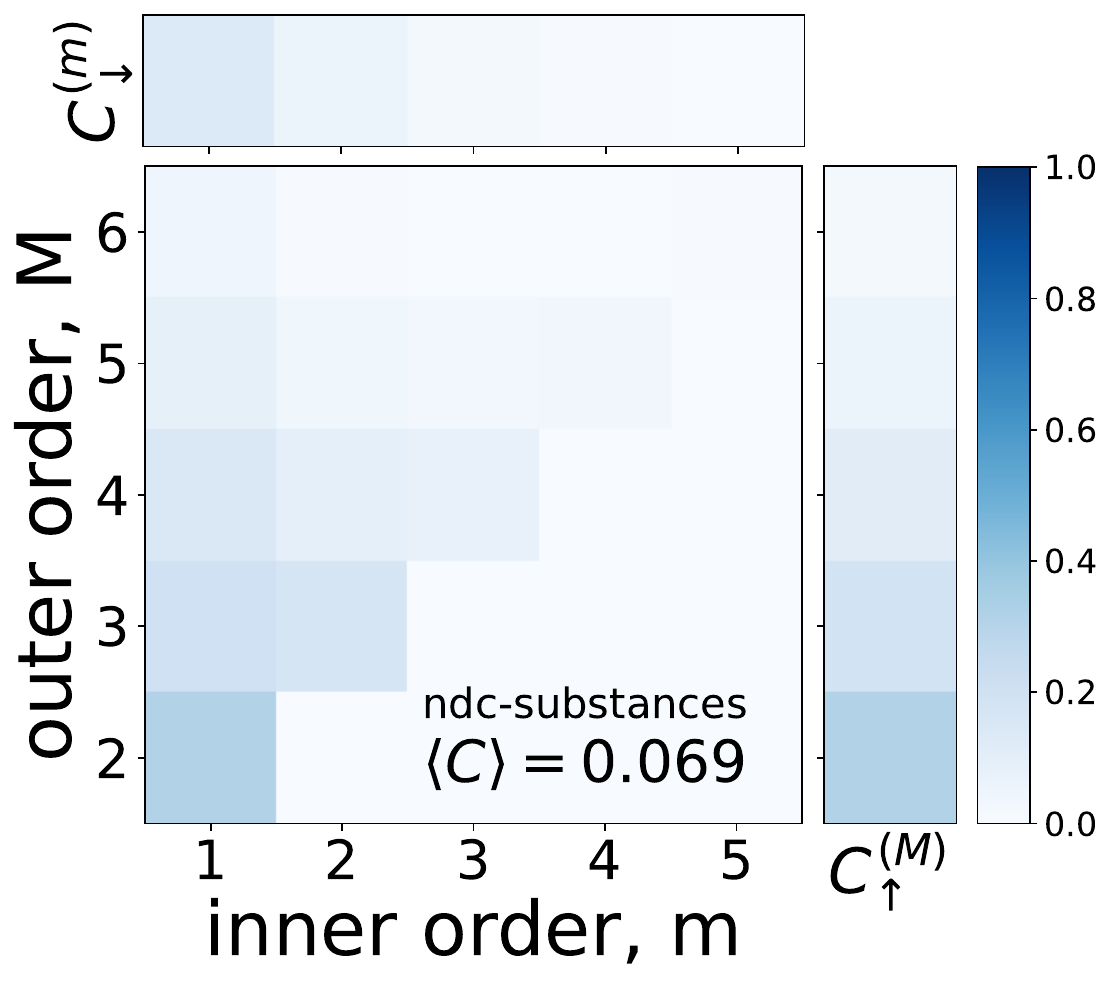}
    \end{subfigure}
    \hfill
    \begin{subfigure}[b]{0.24\textwidth}
        \centering
        \includegraphics[width=\linewidth]{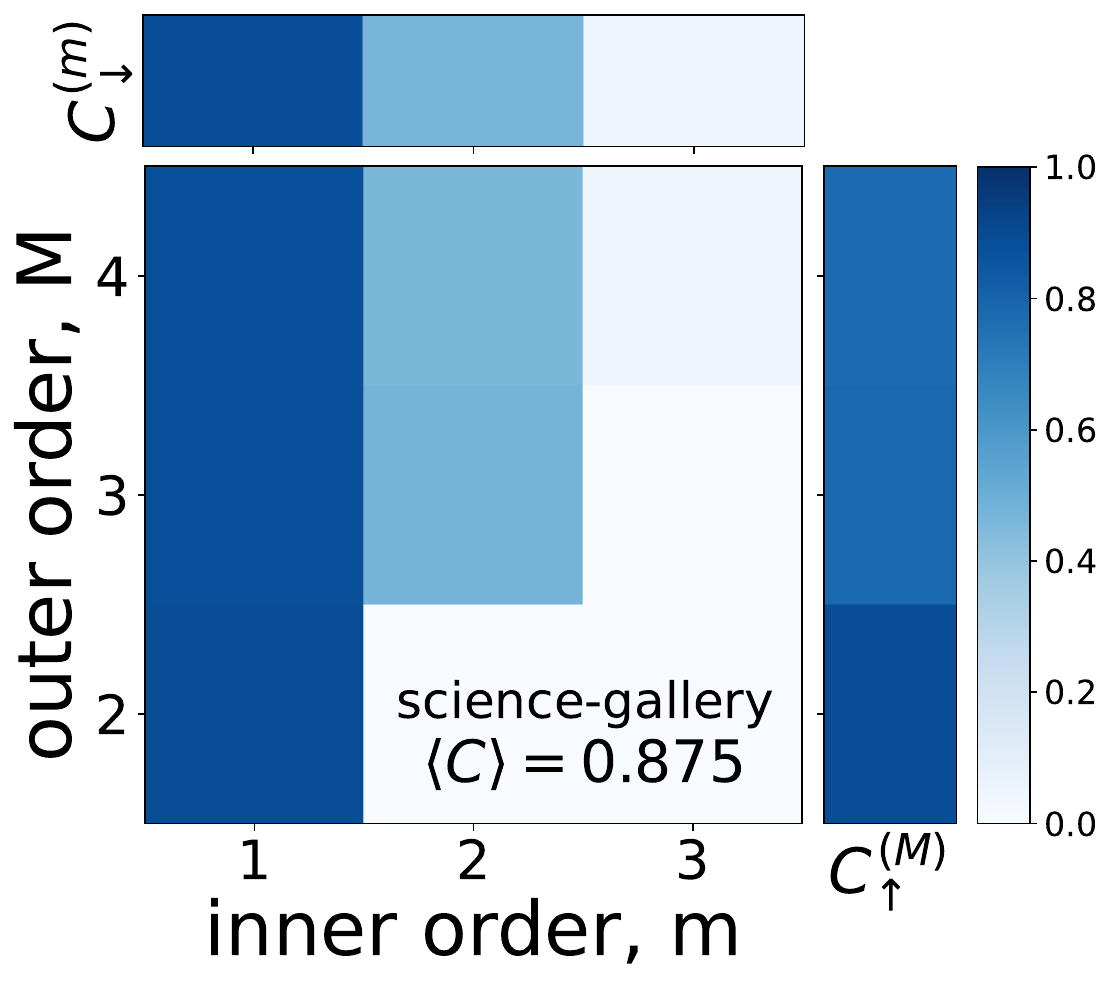}
    \end{subfigure}
    \hfill
    \begin{subfigure}[b]{0.24\textwidth}
        \centering
        \includegraphics[width=\linewidth]{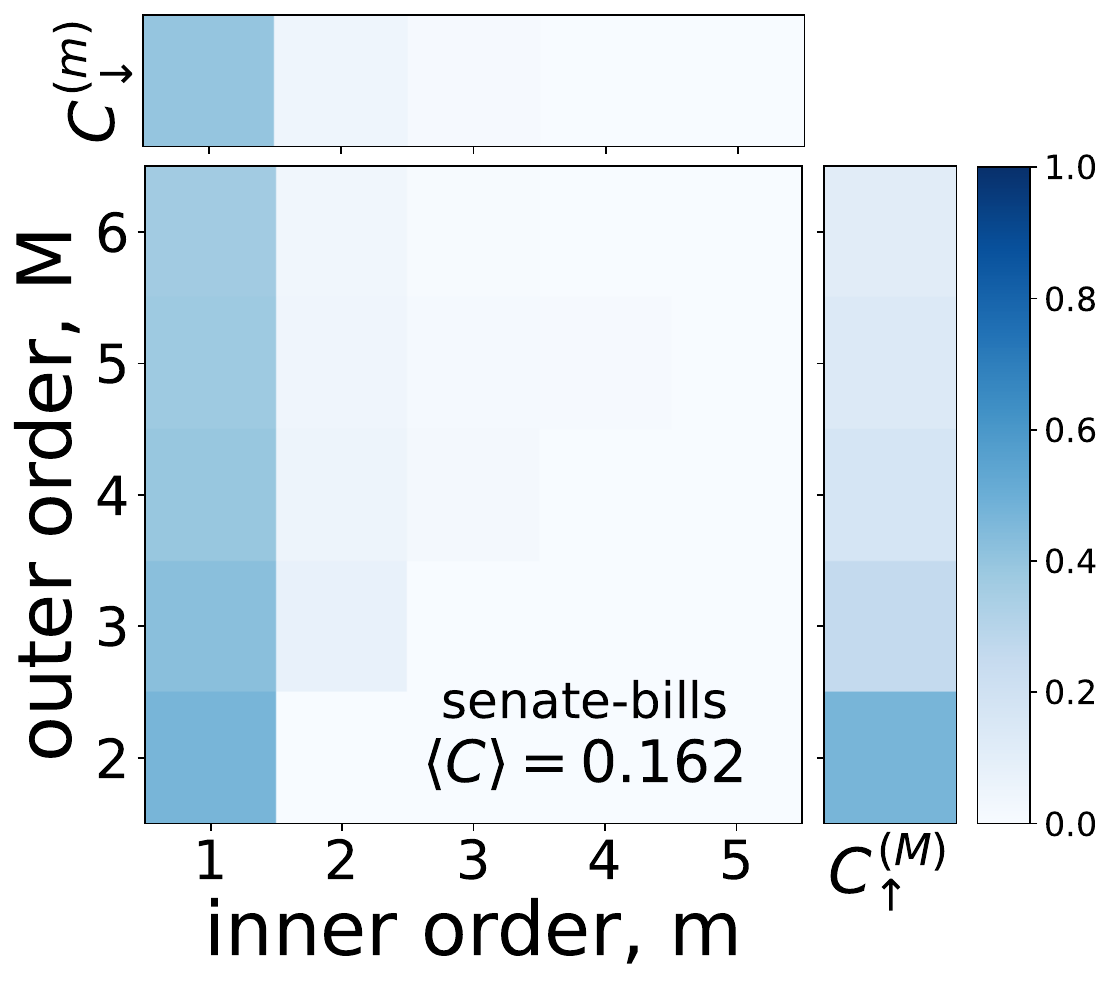}
    \end{subfigure}
    \hfill
    \begin{subfigure}[b]{0.24\textwidth}
        \centering
        \includegraphics[width=\linewidth]{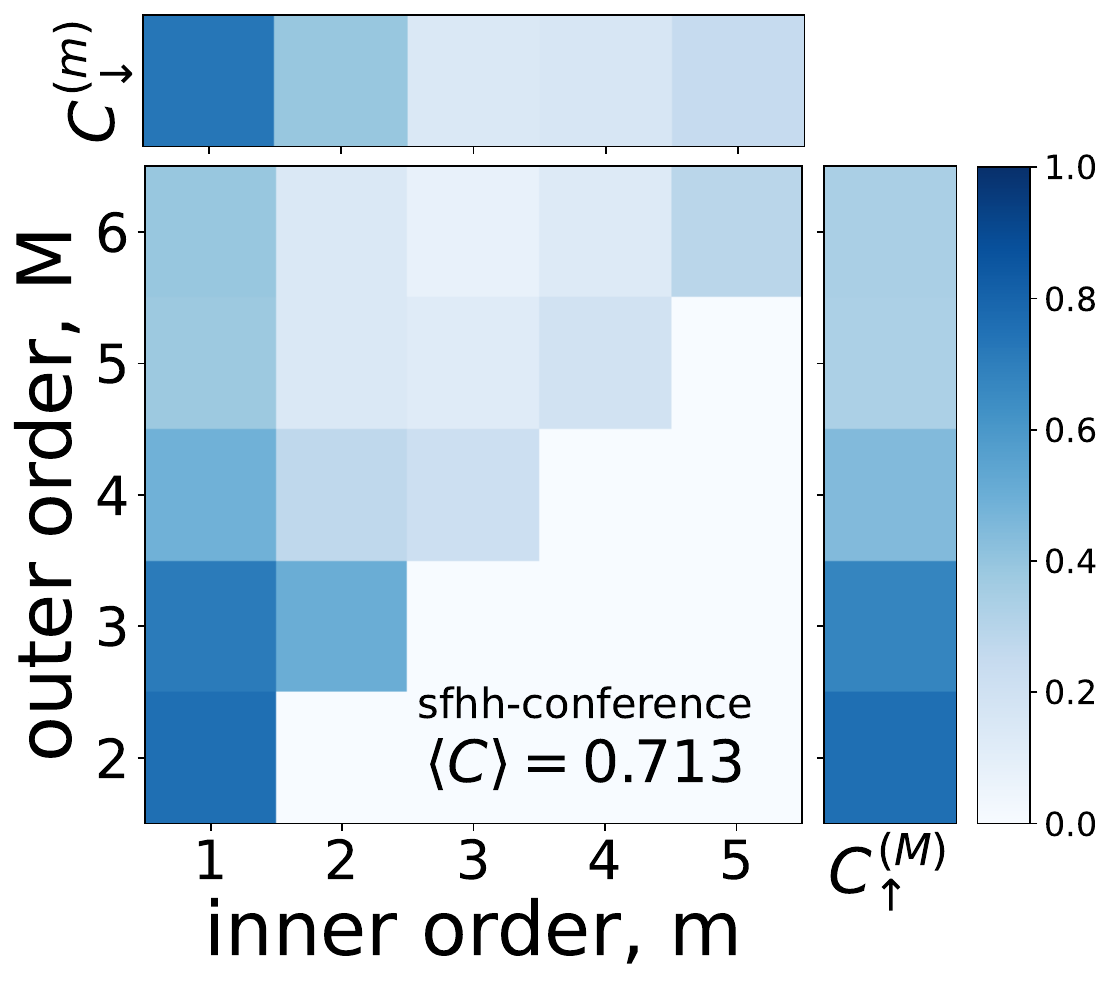}
    \end{subfigure}
    \hfill
    \begin{subfigure}[b]{0.24\textwidth}
        \centering
        \includegraphics[width=\linewidth]{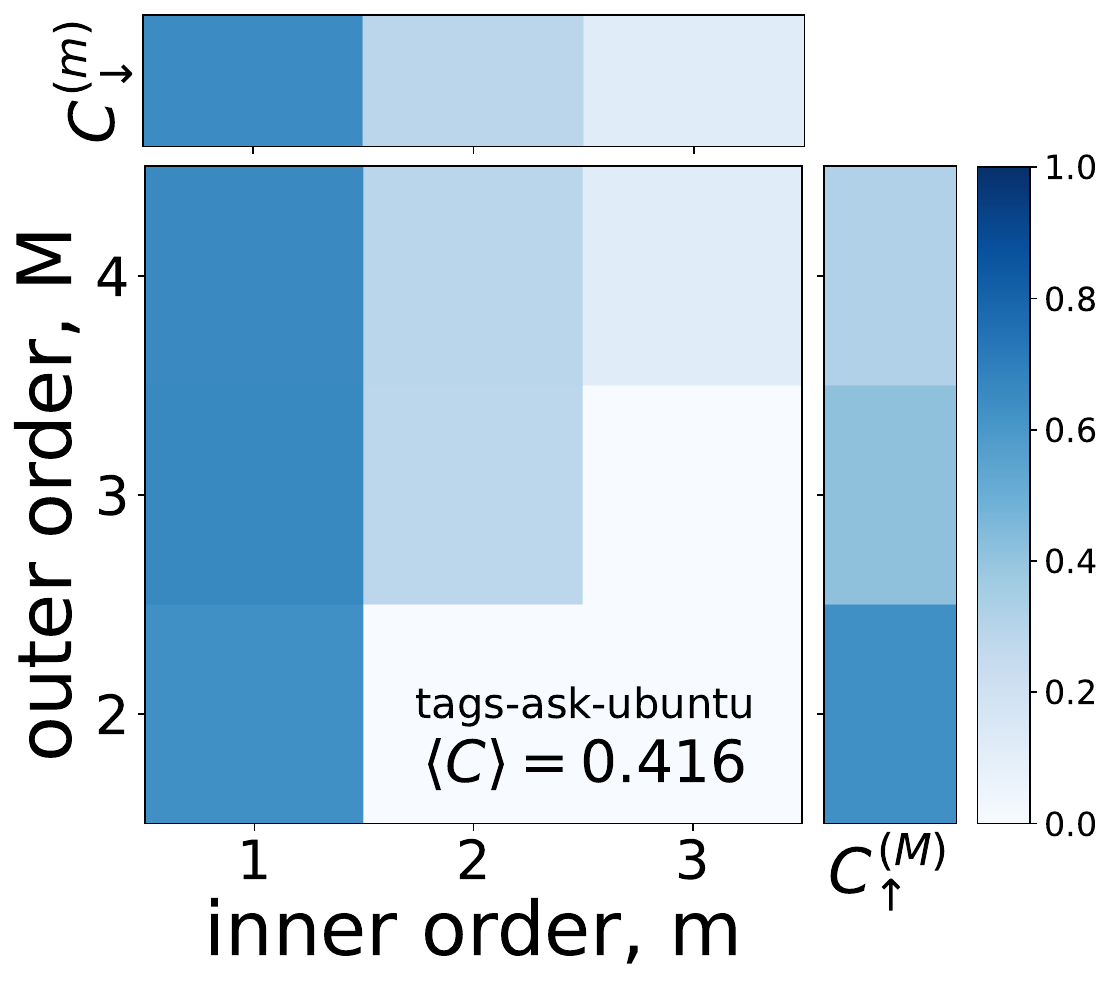}
    \end{subfigure}
    \hfill
    \begin{subfigure}[b]{0.24\textwidth}
        \centering
        \includegraphics[width=\linewidth]{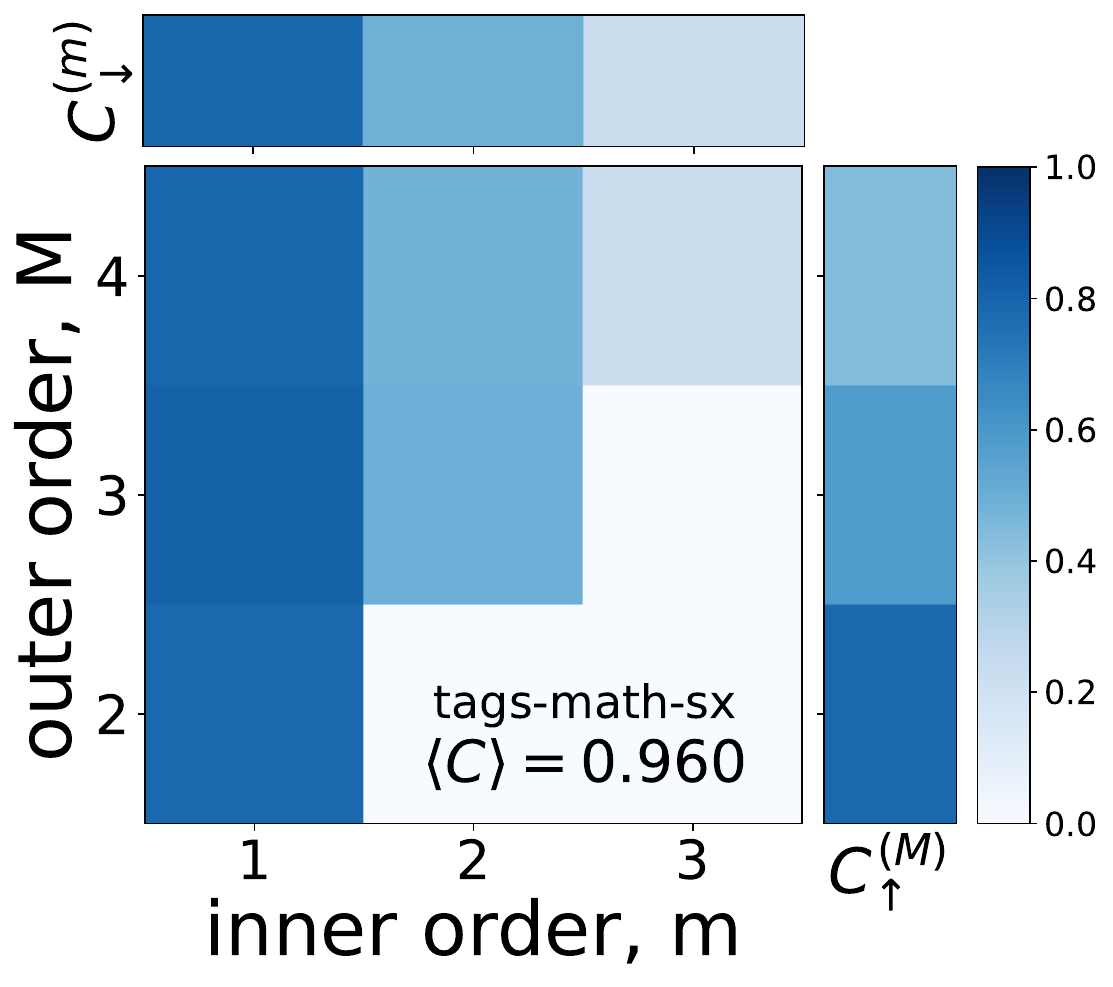}
    \end{subfigure}
    \hfill
    \begin{subfigure}[b]{0.24\textwidth}
        \centering
        \includegraphics[width=\linewidth]{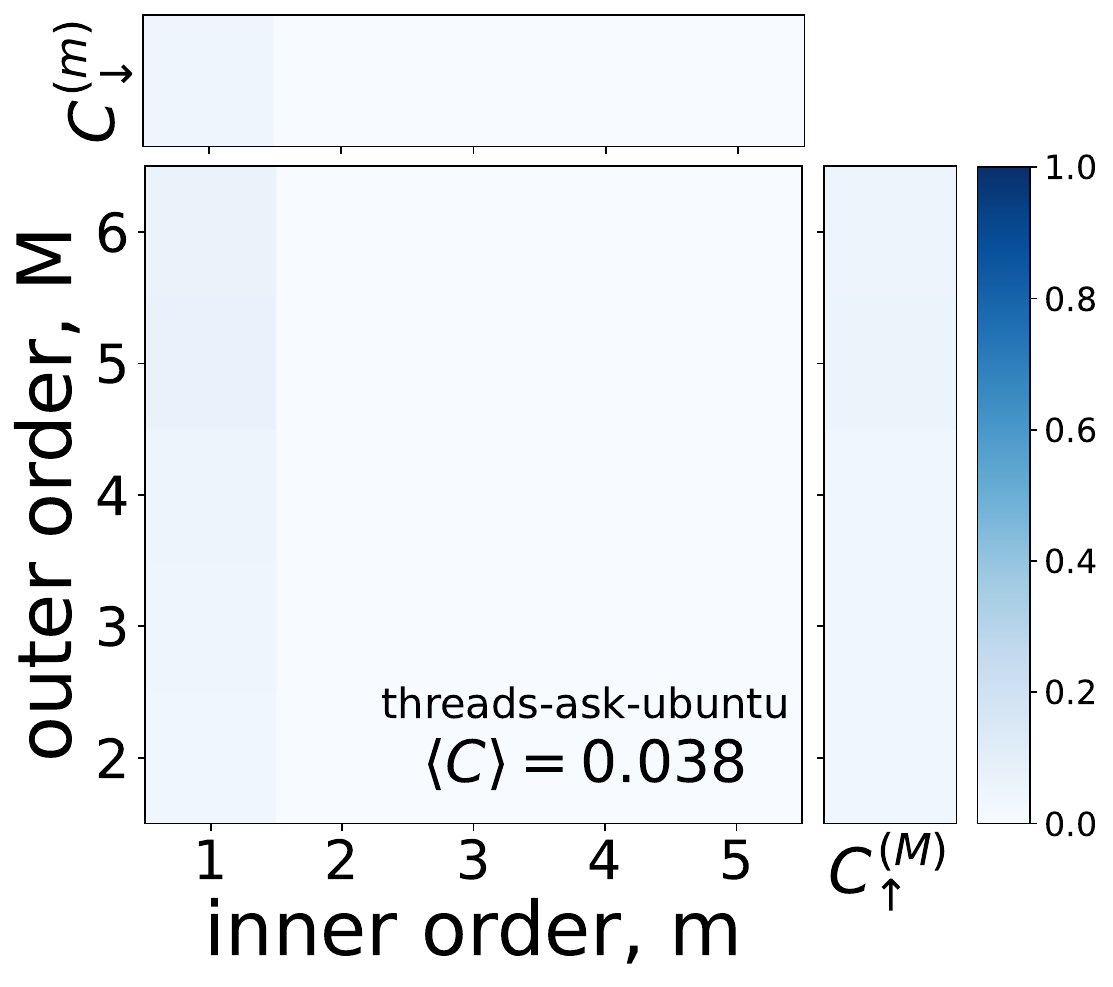}
    \end{subfigure}
    \hfill
    \begin{subfigure}[b]{0.24\textwidth}
        \centering
        \includegraphics[width=\linewidth]{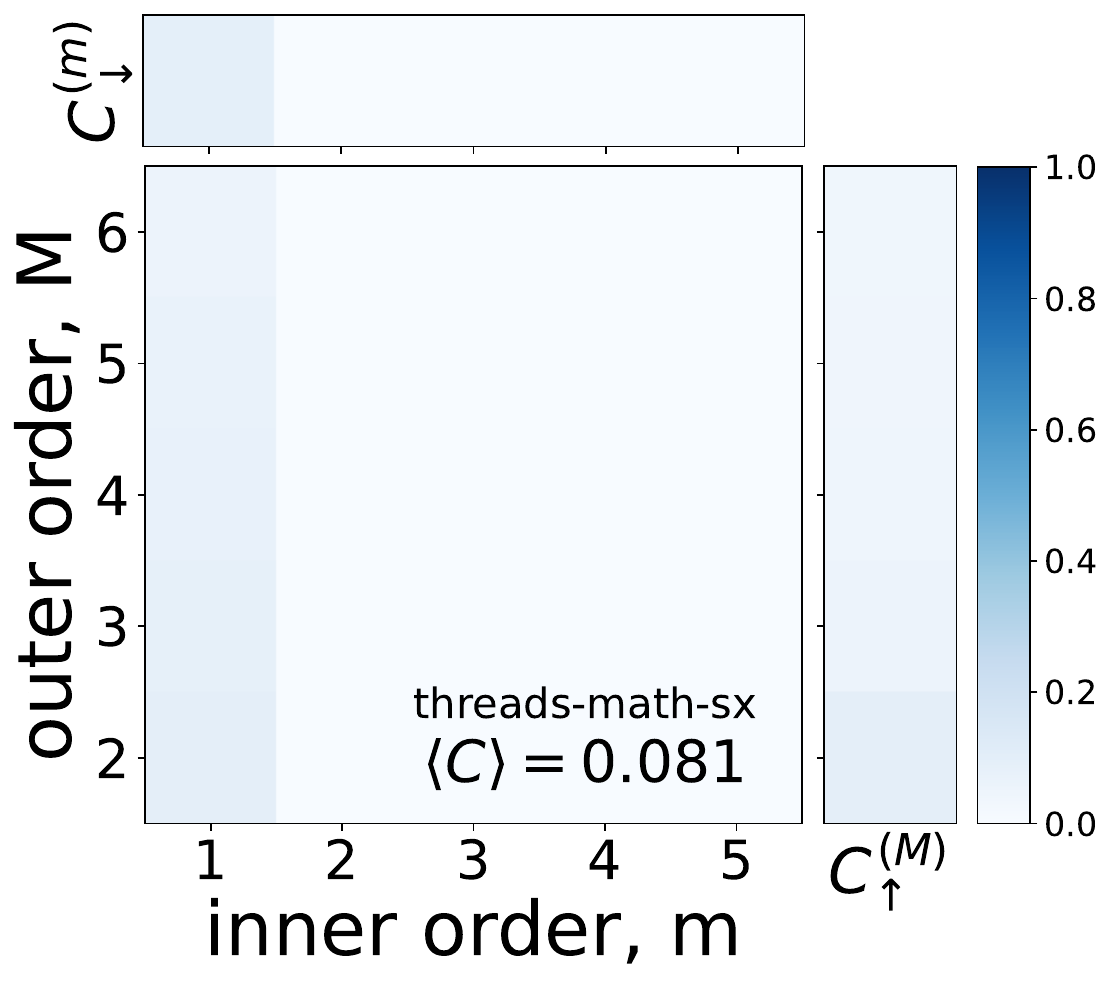}
    \end{subfigure}
    \hfill

    \caption{ Nesting coefficient measured from empirical datasets. Extra rows and columns on top and left represents the average inner and outer-order nesting coefficient respectively. Nesting coefficients for orders greater than 6 were omitted for better visibility. }
    \label{fig:Empirical}
\end{figure*}

\begin{table*}[htb]
    \centering
    \begin{tabular}{c c c c c c c c c c }
        \hline
        Dataset & $N$ & $H$ & $m_{max}$ & $\langle C \rangle$ & $\langle C \rangle_{\text{null}}$ & $\beta^{(1)}_{c,spread}(b=0)$ & $\beta^{(1)}_{c,spread}(b=b_0)$ & $b_0$ & $\mathcal{H}$ \\
        \hline
        congress-bills & 1718 & 104039 & 262 & 0.146 & 0.023 & 0.0238 & 0.0147 $\pm$ 0.0011 & 0.025 & 0.617 $\pm$ 0.046 \\
        contact-high-school & 327 & 7818 & 4 & 0.937 & 0.131 & 0.0300 & 0.0004 $\pm$ 0.0003 & 0.013 & 0.013 $\pm$ 0.010 \\
        contact-primary-school & 242 & 12704 & 4 & 0.941 & 0.278 & 0.0153 & 0.0011 $\pm$ 0.0006 & 0.071 & 0.071 $\pm$ 0.039 \\
        dawn & 2290 & 138742 & 15 & 0.557 & 0.291 & 0.0079 & 0.0046 $\pm$ 0.0008 & 0.003 & 0.582 $\pm$ 0.101 \\
        email-enron & 143 & 1459 & 36 & 0.660 & 0.113 & 0.0796 & 0.0023 $\pm$ 0.0018 & 0.395 & 0.029 $\pm$ 0.022 \\
        email-eu & 986 & 24520 & 39 & 0.675 & 0.105 & 0.0176 & 0.0054 $\pm$ 0.0014 & 0.066 & 0.306 $\pm$ 0.079 \\
        eventernote-events & 69885 & 131647 & 182 & 0.056 & 0.000 & 0.0660 & 0.0557 $\pm$ 0.0050 & 0.138 & 0.844 $\pm$ 0.075 \\ 
        hospital-lyon & 75 & 1824 & 4 & 0.976 & 0.561 & 0.0351 & 0.0008 $\pm$ 0.0002 & 0.137 & 0.022 $\pm$ 0.005 \\
        house-bills & 1494 & 54933 & 261 & 0.067 & 0.011 & 0.0488 & 0.0392 $\pm$ 0.0014 & 0.104 & 0.803 $\pm$ 0.028 \\
        hypertext-conference & 113 & 2434 & 5 & 0.858 & 0.421 & 0.0268 & 0.0026 $\pm$ 0.0018 & 0.459 & 0.097 $\pm$ 0.067 \\
        invs13 & 92 & 787 & 3 & 0.879 & 0.271 & 0.0664 & 0.0025 $\pm$ 0.0015 & 13.30 & 0.037 $\pm$ 0.022 \\
        invs15 & 217 & 4909 & 3 & 0.940 & 0.228 & 0.0260 & 0.0010 $\pm$ 0.0004 & 0.355 & 0.038 $\pm$ 0.015 \\
        malawi-village & 84 & 431 & 3 & 0.982 & 0.150 & 0.1214 & 0.0004 $\pm$ 0.0001 & 2.230 & 0.003 $\pm$ 0.001 \\
        ndc-substances & 3414 & 6417 & 186 & 0.069 & 0.004 & 0.1728 & 0.0826 $\pm$ 0.0241 & 0.465 & 0.478 $\pm$ 0.139 \\
        science-gallery & 410 & 3350 & 4 & 0.875 & 0.048 & 0.0687 & 0.0209 $\pm$ 0.0035 & 0.542 & 0.304 $\pm$ 0.051 \\
        senate-bills & 294 & 21721 & 98 & 0.162 & 0.062 & 0.0345 & 0.0200 $\pm$ 0.0040 & 0.043 & 0.579 $\pm$ 0.115 \\ 
        sfhh-conference & 403 & 10541 & 8 & 0.713 & 0.172 & 0.0198 & 0.0058 $\pm$ 0.0023 & 0.185 & 0.292 $\pm$ 0.116 \\
        tags-ask-ubuntu & 3021 & 145053 & 4 & 0.416 & 0.138 & 0.0110 & 0.0065 $\pm$ 0.0010 & 0.008 & 0.590 $\pm$ 0.091 \\
        tags-math-sx & 1627 & 169259 & 4 & 0.960 & 0.273 & 0.0096 & 0.0007 $\pm$ 0.0002 & 0.043 & 0.072 $\pm$ 0.020 \\
        threads-ask-ubuntu & 82075 & 111802 & 14 & 0.038 & 0.002 & 0.0340 & 0.0235 $\pm$ 0.0051 & 0.277 & 0.691 $\pm$ 0.150 \\
        threads-math-sx & 152702 & 534768 & 21 & 0.081 & 0.003 & 0.0107 & 0.0088 $\pm$ 0.0006 & 0.055 & 0.822 $\pm$ 0.056 \\
        synth-1.00 & 10000 & 56428 & 5 & 1.000 & 0.000 & 0.1820 & 0.0120 & 0.985 & 0.065 \\
        synth-0.75 & 10000 & 56428 & 5 & 0.747 & 0.000 & 0.1820 & 0.0288 & 0.412 & 0.158 \\
        synth-0.50 & 10000 & 56428 & 5 & 0.527 & 0.000 & 0.1820 & 0.0624 & 0.358 & 0.342 \\
        synth-0.25 & 10000 & 56428 & 5 & 0.264 & 0.000 & 0.1820 & 0,1180 & 0.330 & 0.648 \\
        synth-0.00 & 10000 & 56428 & 5 & 0.001 & 0.000 & 0.1820 & 0.1708 & 0.330 & 0.938 \\
        
        \hline
    \end{tabular}
    \caption{Full table with all datasets analyzed and their properties. The overall nesting coefficient of the original network is shown and compared with a null version obtained by rewiring $10 \times H$ pairs of hyperedges. Also shown are the threshold values and higher-order spreading rates $\beta^{(1)}_{c,spread}(b=0)$, $\beta^{(1)}_{c,spread}(b=b_0)$ and $b_0$ used to obtain the hysteresis length $\mathcal{H} = \beta^{(1)}_{c, spread}(b=b_0)/\beta^{(1)}_{c, spread}(b=0)$.}
    \label{tab:Empirical}
\end{table*}

%\bibliography{ref}

\end{document}